\documentclass[3p,times,procedia]{elsarticle}

\usepackage{ecrc}

\volume{00}

\firstpage{1}

\journalname{Procedia Computer Science}

\runauth{}

\jid{procs}

\jnltitlelogo{Procedia Computer Science}

\CopyrightLine{2011}{Published by Elsevier Ltd.}

\usepackage{amscd,amsmath,amsthm,amssymb,amsfonts,bbm}

 \newtheorem{example}{Example}%
\newtheorem{notation}{Notation}
\newtheorem{lemma}{Lemma}
\newtheorem{proposition}{Proposition}

 \newtheorem{definition}{Definition}%

\newcommand{\bigO}{\mathcal{O}}

\newcommand{\pma}{\textup{PerfMatch}}
\newcommand{\RP}{\mathcal{RP}}
\newcommand{\pw}{\textup{pw}}
\newcommand{\twi}{\textup{tw}}
\newcommand{\poly}{\textup{poly}}
\newcommand{\ma}{\textup{Match}}
\newcommand{\RM}{\mathcal{RM}}
\newcommand{\RI}{\mathcal{RI}}
\newcommand{\ind}{\textup{Ind}}

\usepackage[algonl,boxed,norelsize,lined]{algorithm2e}
\usepackage{mathtools}
\usepackage{arydshln}
\usepackage{comment}
\usepackage[all]{xy}
\usepackage{datetime}
\usepackage{subfig}
\usepackage{color,mathtools}
\usepackage[T1]{fontenc}
\usepackage{hyperref}
\usepackage{graphicx}
\usepackage[new]{old-arrows}
\hypersetup{
	colorlinks = true,
	linkcolor = blue,
	anchorcolor = blue,
	citecolor = teal,
	filecolor = blue,
	urlcolor = black
}
\usepackage{tocloft}

\usepackage{booktabs}
\usepackage{paralist}
\setdefaultleftmargin{0pt}{0pt}{0pt}{0pt}{0pt}{0pt}

\usepackage{wrapfig}

\usepackage{amsmath}
\usepackage{amssymb}
\usepackage{amsthm}
\usepackage{verbatim}
\usepackage{graphicx}
\usepackage[disable]{todonotes}
\usepackage{floatrow}

\usepackage{tikz,tikz-cd}
\usetikzlibrary{decorations.pathreplacing}
\usetikzlibrary{shapes.multipart}
\usetikzlibrary{calc,cd}

\usepackage{float}

\renewcommand{\paragraph}[1]{\smallskip\noindent\textbf{#1.}}

\usepackage{graphicx}
\usepackage{amsmath}
\usepackage{xcolor}
\usepackage{hyperref}
\usepackage{nicefrac}
\usepackage{mathtools}
\usepackage{caption}
\usepackage{wrapfig}
\usepackage{chemfig}
\DeclarePairedDelimiter\abs{\lvert}{\rvert}

\usepackage{paralist}
\usepackage{appendix}

\usepackage{amssymb}

\usepackage[figuresright]{rotating}

\begin{document}

\begin{frontmatter}



\dochead{XIII Latin American Algorithms, Graphs, and Optimization Symposium}

\title{Combinatorial Parameterized Algorithms for Chemical Descriptors based on Molecular Graph Sparsity}

 \author[hkust]{Giovanna Kobus Conrado}
 \author[oxford]{Amir Kafshdar Goharshady}
 \author[mpi]{Harshit Jitendra Motwani}
 \author[hkust]{Sergei Novozhilov}
 \address[hkust]{Hong Kong University of Science and Technology, Clear Water Bay, Hong Kong\\\texttt{\{gkc, snovozhilov\}@connect.ust.hk}}
 \address[oxford]{University of Oxford, Oxford, United Kingdom \\\texttt{amir.goharshady@cs.ox.ac.uk}}
 \address[mpi]{Max Planck Institute for Software Systems, Kaiserslautern, Germany\\\texttt{hmotwani@mpi-sws.org}}

\begin{abstract}
We present efficient combinatorial parameterized algorithms for several classical graph-based counting problems in computational chemistry, including (i)~Kekulé structures, (ii)~the Hosoya index, (iii)~the Merrifield–Simmons index, and (iv)~Graph entropy based on matchings and independent sets. All these problems were known to be $\# P$-complete. Building on the intuition that molecular graphs are often sparse and tree-like, we provide fixed-parameter tractable  (FPT) algorithms using treewidth as our parameter. We also provide extensive experimental results over the entire PubChem database of chemical compounds, containing more than 113 million real-world molecules. In our experiments, we observe that the molecules are indeed sparse and tree-like, with more than $99.9\%$ of them having a treewidth of at most $5.$ This justifies our choice of parameter. Our experiments also illustrate considerable improvements over the previous approaches. Based on these results, we argue that parameterized algorithms, especially based on treewidth, should be adopted as the default approach for problems in computational chemistry that are defined over molecular graphs. 
\end{abstract}


\end{frontmatter}


\section{Introduction}\label{sec:intro}

\paragraph{Molecular Descriptors and Topological Indices} In computational chemistry, a molecular descriptor is simply a function that maps graphs modeling molecules to numbers. An important family of molecular descriptors are the topological indices, i.e.~numerical graph invariants that characterize the topology of graphs associated with molecules~\cite{trinajstic2018chemical}. They have been studied for more than a century and hundreds of them are available and used in the literature~\cite{todeschini2008handbook}. These indices play a vital role in computational chemistry as they are used as molecular descriptors for QSAR (Quantitative Structure-Activity Relationship) and QSPR (Quantitative Structure-Property Relationship) \cite{yousefinejad2015chemometrics,dearden2017use}. In other words, the topology and geometrical properties of hydrogen-suppressed graphs of molecules can be used to describe their physical and chemical properties \cite{bonchev2018chemical}. This in turn helps in predicting the biological activity, toxicity, and other properties of molecules. Topological indices provide a single numerical value that can describe otherwise complicated and intricate topological and geometrical properties of molecules. Due to their efficacy in capturing molecular properties, topological indices are often used for drug design, toxicity detection, and many other applications in computational chemistry~\cite{nantasenamat2010advances,ghasemi2018neural}.

\paragraph{Connectivity and Distance-based Indices} The many topological indices that have been proposed and studied in the literature~\cite{todeschini2008handbook} can be broadly classified in two categories: connectivity-based and distance-based \cite{trinajstic2018chemical, todeschini2008handbook, leszczynski2012handbook, xue2000molecular}. The former are based on the adjacency relationships among the vertices of the graph, while the latter use the distances. Some well-known connectivity-based indices are the so-called ``Zagreb group'' indices~\cite{nikolic2003zagreb}, the Randi{\'c} index~\cite{randic2001connectivity} and the Hosoya index~\cite{hosoya1971topological}. On the other hand, classical distance-based indices include the Wiener index~\cite{nikolic1995wiener}, the Balaban index~\cite{balaban2000historical} and information-theoretic indices~\cite{basak2000information}.

\paragraph{Treewidth~\cite{robertson1984graph}}
Treewidth is a well-studied parameter for graphs~\cite{cygan2015parameterized}. Informally, it is a measure of tree-likeness of a graph~\cite{bodlaender1993tourist}. Trees and forests have a treewidth of $1$ and graphs with treewidth $k$ can be decomposed into small parts, called ``bags'', of size at most $k+1$ which are in turn connected to each other in a tree-like manner. Many NP-hard graph problems are known to have efficient solutions when restricted to graphs with bounded treewidth~\cite{DBLP:conf/icalp/Bodlaender88}. See Section~\ref{sec:prelim} for a formal definition.


\paragraph{Our Focus} Given their significance in computational chemistry and biology, computing chemical descriptors is a natural algorithmic problem. Some of these indices are computable in polynomial time. Unfortunately, for many other well-studied studied topological indices, such as the ones below, the corresponding counting problems are known to be $\# P$-complete~\cite{valiant1979complexity}, leading to best-known algorithms that can only scale to molecules with a tiny number of atoms. In this work, we focus on developing fixed-parameter tractable (FPT) algorithms using treewidth as the parameter. This is motivated by the observation that real-world molecules are sparse, due to the limited valency of atoms, and often have a tree-like appearance. Thus, we guess that they should have small treewidth. In particular, we focus on computing the following:
\begin{compactitem}
	\item \textbf{Counting Kekulé Structures:}  A Kekulé structure of a molecule is essentially a perfect matching of the underlying graph. Therefore, counting Kekulé structures is equivalent to finding the total number of perfect matchings of the underlying graph~\cite{trinajstic2018chemical}.
	\item \textbf{Hosoya Index:} The Hosoya index, also known as the Z index, of a graph is the total number of matchings of the graph~\cite{hosoya1971topological}. 
	\item \textbf{Merrifield–Simmons Index:} The Merrifield–Simmons index of a graph is the total number of independent sets of the graph~\cite{merrifield1980structures}.
	\item \textbf{Graph Entropy:} Graph Entropy is defined based on the number of matchings and independent sets of every possible size in a given graph~\cite{cao2017network}. Specifically, for a graph $G$ with $n$ vertices and $m$ edges, we have
	\begin{equation} \textstyle \label{eq:entropy}
	I_v(G) := - \sum_{k=0}^n \frac{i_k(G)}{i(G)} \cdot \log \frac{i_k(G)}{i(G)}
	~~~\text{ and }~~~
	I_e(G) := -\sum_{k=0}^m \frac{i'_k(G)}{i'(G)} \cdot \log \frac{i'_k(G)}{i'(G)}.
\end{equation}
	Here, $I_v(G)$ is the \emph{graph entropy based on independent sets}, $i_k(G)$ is the number of independent sets of size $k$ in the graph $G$ and $i(G)$ is the total number of independent sets, i.e.~$i(G) = \sum_{k=0}^n i_k(G).$ Similarly, $I_e(G)$ is the \emph{graph entropy based on matchings} and $i'_k(G)$ is the number of matchings with $k$ edges in $G$ with $i'(G) := \sum_{k=0}^m i'_k(G)$~\cite{wan2018computing}.
\end{compactitem}

It is well-known that the first three problems above are $\# P$-complete~\cite{valiant1979complexity,jerrum1987two,dyer2000markov}. Currently, the existing non-parameterized approaches take exponential time in the worst case and do not scale up for molecules with a large number of atoms and bonds. There are known FPT algorithms parameterized by the treewidth~\cite{wan2018computing} in the literature. However, we show that one can improve these algorithms significantly, as well.


\paragraph{Our Contribution} On the theoretical side, we present several algorithms for the aforementioned classical problems in computational chemistry. Our algorithms are fixed-parameter tractable (FPT) and run in polynomial time for graphs with bounded treewidth. They are simple and follow the paradigm of dynamic programming over tree decompositions. Moreover, they significantly improve the previous asymptotic runtime for combinatorial solutions to these problems as summarized in Table~\ref{tab:runtime}.
On the practical side, we provide extensive experimental results over more than 113 million real-world molecules from the PubChem database~\cite{kim2023pubchem}. Our experiments demonstrate that more than $99.9\%$ of molecules in the PubChem database have a treewidth of $5$ or smaller.  Thus, our algorithms are directly applicable to them. 
Based on our results, we believe developing parameterized algorithms and specifically using treewidth as the parameter, should be adopted as the default strategy when dealing with graph problems in computational chemistry. Finally, we also experimentally compare the runtime of our algorithms with previous methods in the literature, over the entire PubChem database, showing that the theoretical advances lead to significant runtime improvements in practice, as well.

\paragraph{Related Parameterized Results} Counting all matchings or independent sets of a given graph is a problem expressible in the monadic second-order logic and thus FPT with respect to treewidth, based on Courcelle's well-known theorem~\cite{courcelle2001fixed}. However, algorithms obtained by Courcelle's theorem are not practical and this approach is also not applicable to counting matchings/independent sets of a desired size. The work~\cite{wan2018computing} provides parameterized algorithms to find the number of independent sets or matchings of any desired size using treewidth as the parameter. This can directly be applied to our setting. In comparison to~\cite{wan2018computing}, our approach is more efficient by an almost-linear factor for graph entropies and a quadratic factor for the indices. We also improve the runtime's dependence on treewidth. See Table~\ref{tab:runtime}. Our algorithms are not asymptotically optimal with respect to the parameter. Specifically, the runtime's dependence on treewidth can be further improved using techniques that employ fast subset convolution~\cite{DBLP:journals/corr/abs-1806-01667}. However, there are two downsides to this: (i)~some convolution-based algorithms produce incorrect results in practice due to floating-point precision errors, and (ii)~the runtime dependence on $n$ increases by a logarithmic factor. The former issue is of course unacceptable in our setting, thus we design purely combinatorial algorithms that avoid floating-point computations. As for the latter issue, we will see in Section~\ref{sec:Experiments} that 99.9\% of molecules have a treewidth of less than $5.$ Thus, the primary goal is to reduce the runtime's dependence on $n,$ rather than $\twi.$

\begin{table}[H]
	\centering
	\small 
	\begin{tabular}{llccc}
	\toprule
	  \textbf{Counting Problem} & \textbf{WTZL~\cite{wan2018computing}} & \textbf{Our Algorithm} \\
	\midrule
	Kekulé  & $\bigO(n^3 \cdot \textup{poly}(\twi) \cdot 4^{\twi})$ & $\bigO(n\cdot \textup{poly}(\twi)\cdot3^{\twi})$ \\
	Hosoya & $\bigO(n^3 \cdot \textup{poly}(\twi) \cdot 4^{\twi})$ & $\bigO(n\cdot \textup{poly}(\twi)\cdot3^{\twi})$ \\
	Merrifield--Simmons &  $\bigO(n^3 \cdot \textup{poly}(\twi) \cdot 2^{\twi})$ & $\bigO(n\cdot \textup{poly}(\twi)\cdot2^{\twi})$ \\
	Matchings (all sizes) &  $\bigO(n^3 \cdot \twi \cdot 4^\twi)$ & $\bigO(n^2 \cdot \log n  \cdot \poly(\twi)\cdot3^{\twi})$ \\
	Independent Sets (all sizes) &  $\bigO(n^3 \cdot \twi \cdot 2^\twi)$ & $\bigO(n^2 \cdot \log n  \cdot \poly(\twi)\cdot2^{\twi})$ \\
	\bottomrule
	\end{tabular}
	\caption{Runtime comparison of our algorithms and those of~\cite{wan2018computing}. Here, $n$ is the number of vertices, $m$ is the number of edges, and $\twi$ is the treewidth of the graph.}
	\label{tab:runtime}
\end{table}


\section{Preliminaries}\label{sec:prelim}



\begin{definition}[Path Decomposition \cite{DBLP:journals/jct/RobertsonS83,cygan2015parameterized}]\label{def:path_dec}
A \textit{path decomposition} of a graph $G = (V, E)$ is a sequence $\mathcal{P} = \{X_1, \dots, X_r\}$ of ``bags'', where each bag $X_i$ is a subset of $V,$ such that following conditions hold:
\begin{compactenum}
	\item For each $v \in V$, there exists a pair of indices 
	$1 \leq l(v) \leq r(v) \leq r$ such that $v \in X_i \Leftrightarrow l(v) \leq i \leq r(v),$ i.e.~each vertex of the graph $G$ appears in a contiguous segment of bags.  
	\item For each $uv \in E$, there exists an index $i$ such that $\{u, v \} \subseteq X_i,$ i.e.~there is a bag that contains both endpoints of the edge.
\end{compactenum}

\begin{definition}[Pathwidth~\cite{DBLP:journals/jct/RobertsonS83}]\label{def:pathwidth}
The \emph{width} of a path decomposition $\mathcal{P} = \{X_1, \dots, X_r\}$ is the size of its largest bag minus one, i.e.~$\max_{1 \leq i \leq r} \lvert X_i \rvert - 1$. The \emph{pathwidth} of a graph $G$, denoted by $\pw(G)$, is the minimum possible width among path decompositions of $G.$
\end{definition}

When designing algorithms, it is often useful to turn decompositions into the following folklore form:

\end{definition}
\begin{definition}[Nice Path Decomposition]\label{def:nice_path_dec}
A \emph{path decomposition} $\mathcal{P} = \{X_1, \dots, X_r\}$ is nice if it satisfies the following additional constraints:
\begin{compactenum}
	\item $X_1 = X_r = \emptyset$.
	\item For every $i \geq 1,$ the bag $X_{i+1}$ is of one of the following types:
	\begin{compactitem}
		\item \emph{Forget Node:} There exists a vertex $v \in X_i$ such that $X_{i + 1} = X_i \setminus \{v\}$. In this case, we say that $X_{i+1}$ \emph{forgets} $v.$
		\item  \emph{Introduce Node}: There exists a vertex $v \in V \setminus X_i$ such that $X_{i + 1} = X_i \cup \{v\}$. We say that $X_{i+1}$ \emph{introduces} $v.$
	\end{compactitem}
\end{compactenum}
It is well-known that every path decomposition can be turned into a nice decomposition of the same width in linear time~\cite{cygan2015parameterized}.
\end{definition}

\begin{definition}[Tree Decomposition \cite{robertson1984graph,cygan2015parameterized}]\label{def:tree_dec} 
A \textit{tree decomposition} of a graph $G$ is a pair $\mathcal{T} = (T, \{X_t\}_{t \in V(T)})$, where $T$ is a rooted tree with root $r$, each bag $X_t$ is a subset of vertices of $G$ and the following conditions hold:
\begin{compactenum}
	\item For every $uv \in E(G)$, there exists a node $t \in V(T)$ such that $\{u, v\} \subseteq X_t$. In other words, every edge is covered by some bag.
	\item For every $v \in V(G)$, the set $T_v := \{t \in V(T): v \in X_t\}$, consisting of all nodes of the tree whose bags contain $v,$ forms a non-empty and connected subtree of $T.$ In other words, every vertex is covered by some bag and the set of bags covering each vertex is a subtree of $T.$
\end{compactenum}
\end{definition}

\begin{definition}[Treewidth \cite{robertson1984graph}]\label{def:treewidth}
	The \emph{width} of a tree decomposition $\mathcal{T} = (T, \{X_t\}_{t \in V(T)})$ is defined as $\max_{t \in V(T)} \lvert X_t \rvert - 1.$ The \emph{treewidth} of a graph $G$, denoted by $\twi(G)$, is the minimum possible width among tree decompositions of $G$.
\end{definition}

Given that every path decomposition is by definition a tree decomposition, too, we always have $\pw(G) \geq \twi(G).$ We consider the last bag $r$ of a path decomposition as its root. Moreover, we can define an analogous notion of niceness for tree decompositions:

\begin{definition}[Nice Tree Decomposition \cite{cygan2015parameterized}]\label{def:nice_tree_dec}
The tree decomposition $\mathcal{T} = (T, \{X_t\})$ is \emph{nice} if it satisfies the following conditions:
\begin{compactenum}
	\item The root bag is empty, i.e.~$X_r = \emptyset.$
	\item If $l$ is a leaf of the $T$, then $X_l = \emptyset.$
	\item Each non-leaf node of tree $T$ is of one of the following three types: 
	\begin{compactitem}
		\item \emph{Forget Node:} If $b$ is a forget node, it has exactly one child $c$ and there is a vertex $v \in X_c$ such that $X_b = X_c \setminus \{v\}$. We say that $b$ forgets $v.$
		\item \emph{Introduce Node:} If $b$ is an introduce node, it has exactly one child $c$ and there is a vertex $v \in V(G) \setminus X_c$ such that $X_b = X_c \cup \{v\}$. We say that $b$ introduces $v.$
		\item \emph{Join Node:} If $b$ is a join node, it has exactly two children ${c_1}$ and ${c_2}$ such that $X_b = X_{c_1} = X_{c_2}.$ 
		\end{compactitem}
\end{compactenum}
It is well-known that every tree decomposition can be turned into a nice tree decomposition of the same width in linear time~\cite{cygan2015parameterized}.
\end{definition}


\begin{notation}\label{notation:subtree}
We write $T_t$ to denote the subtree of $T$ rooted at $t.$ We also define $G^\downarrow_t := G\left[\cup \{X_t: t \in T_t\}\right].$ In other words, $G^\downarrow_t$ is the subgraph of $G$ induced on vertices that appear in the bags at $t$ or its descendants.
\end{notation}


\begin{example}\label{example:explaining_tree_decomposition}
	Figure~\ref{fig:tree_dec_caffeine} (left) is the caffeine molecule. Figure~\ref{fig:tree_dec_caffeine} (center) is a graph representation of the same molecule and Figure~\ref{fig:tree_dec_caffeine} (right) is a tree decomposition of this graph with width $2.$ This is an optimal decomposition and thus the treewidth of caffeine is $2.$

	\begin{figure}
		\centering
		\subfloat[][Caffeine]{\includegraphics{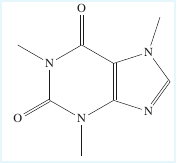}}
		\quad
		\subfloat[][Caffeine's Graph]{\resizebox{0.22\textwidth}{!}{
\begin{tikzpicture}[node distance={12mm}, line width=1pt, main/.style = {draw, circle}] 
\node[main,fill=cyan!20] (1) []{$C_1$}; 
\node[main, fill=green!20, dashed] (2) [below right of=1]{$C_2$}; 
\node[main, fill=green!20, dashed] (3) [below of=2]{$C_3$}; 
\node[main,fill=cyan!20] (4) [below left of=3]{$N_1$}; 
\node[main,fill=cyan!20] (5) [above left of=4]{$C_4$}; 
\node[main,fill=cyan!20] (6) [above of=5]{$N_2$};
\node[main,fill=orange!20] (9) [below right of=3]{$N_4$}; 
\node[main,fill=orange!20] (8) [above right of=9]{$C_5$}; 
\node[main, fill=orange!20] (7) [above of=8]{$N_3$};
\node[main,fill=cyan!20] (10) [above of=1]{$O_1$}; 
\node[main,fill=cyan!20] (11) [above left of=4]{$C_4$}; 
\node[main,fill=cyan!20] (12) [above of=5]{$N_2$}; 

\node[main,fill=cyan!20] (13) [below of=4]{$C_6$};  

\node[main,fill=cyan!20] (17) [below left of=5]{$O_2$};

\node[main,fill=cyan!20] (18) [above left of=6]{$C_7$}; 

\node[main,fill=orange!20] (22) [above right of=7]{$C_8$};

\draw [] (1) -- (2); 
\draw [] (2) -- (3); 
\draw [] (3) -- (4); 
\draw [] (4) -- (5); 
\draw [] (5) -- (6); 
\draw [] (6) -- (1); 
\draw [] (2) -- (7); 
\draw [] (7) -- (8); 
\draw [] (8) -- (9); 
\draw [] (9) -- (3); 
\draw [] (1) -- (10); 

\draw [] (4) -- (13); 

\draw [] (5) -- (17); 

\draw [] (6) -- (18); 

\draw [] (7) -- (22); 
\end{tikzpicture}
}}
		\quad
		\subfloat[][A Path Decomposition]{\resizebox{0.16\textwidth}{!}{
\begin{tikzpicture}[node distance={7mm}, thick, main/.style = {draw, rectangle}] 
\node[main, fill=green!20, dashed] at (0,0) (8) {$C_2,C_3$};
\node[main,fill=cyan!20] at (-1,-1) (7){$C_1,C_2,C_3$};
\node[main,fill=cyan!20] (100) [below of =7] {$C_1,C_3,N_1$};
\node[main,fill=cyan!20] (6) [below of =100]{$C_1,N_1,C_2$};
\node[main,fill=cyan!20] (5) [below of =6]{$C_1,N_1,C_6$};
\node[main,fill=cyan!20] (4) [below of =5]{$C_4,C_1,N_1$};
\node[main,fill=cyan!20] (3) [below of =4]{$C_4,C_1,O_1$};
\node[main,fill=cyan!20] (2) [below of =3]{$C_4,C_1,O_2$};
\node[main,fill=cyan!20] (1) [below of =2]{$N_2,C_4,C_1$};
\node[main,fill=cyan!20] (0) [below of =1] {$N_2,C_7$}; 


\node[main,fill=orange!20] at (1, -1) (9){$C_2,C_3,N_4$};
\node[main,fill=orange!20] (10) [below of =9]{$C_2,N_4,N_3$};
\node[main,fill=orange!20] (11) [below of =10]{$N_4,N_3,C_5$};
\node[main,fill=orange!20] (12) [below of =11]{$N_3, C_8$};

\draw [] (0) -- (1); 
\draw [] (1) -- (2); 
\draw [] (2) -- (3); 
\draw [] (3) -- (4); 
\draw [] (4) -- (5); 
\draw [] (5) -- (6); 
\draw [] (6) -- (100); 
\draw [] (7) -- (100); 
\draw [] (7) -- (8); 
\draw [] (8) -- (9); 
\draw [] (9) -- (10); 
\draw [] (10) -- (11); 
\draw [] (11) -- (12);

\end{tikzpicture}
}}
		\quad
		\subfloat[][A Tree Decomposition]{\resizebox{0.3\textwidth}{!}{
\begin{tikzpicture}[node distance={15mm}, thick, main/.style = {draw, rectangle}] 
\node[main, fill=green!20, dashed] (0) []{$C_2,C_3$}; 
\node[main,fill=cyan!20] (00) [below left of =0]{$C_2,C_3,N_1$};
\node[main,fill=cyan!20] (000) [below left of =00]{$N_1,C_6$};
\node[main,fill=cyan!20] (001) [below right of =00]{$C_1,C_2,N_1$};
\node[main,fill=cyan!20,fill=cyan!20] (0011) [below right of =001]{$C_1,O_1$};
\node[main,fill=cyan!20] (0010) [below left of =001]{$C_1,C_4,N_1$};
\node[main,fill=cyan!20] (00101) [below right of =0010]{$C_4,O_2$};
\node[main,fill=cyan!20] (00100) [below left of =0010]{$C_1,C_4,N_2$};
\node[main,fill=cyan!20] (001000) [below of =00100,,node distance={10mm}]{$N_2,C_7$};
\node[main,fill=orange!20] (01) [below right of =0]{$C_2,C_3,N_4$};
\node[main,fill=orange!20] (011) [below right of =01]{$C_2,N_4,C_5$};
\node[main,fill=orange!20] (0111) [below right of =011]{$C_2,N_3,C_5$};
\node[main,fill=orange!20] (01110) [below of =0111,,node distance={10mm}]{$N_3,C_8$};

\draw [] (0) -- (00); 
\draw [] (00) -- (000); 
\draw [] (00) -- (001); 
\draw [] (001) -- (0010); 
\draw [] (001) -- (0011); 
\draw [] (0010) -- (00100); 
\draw [] (00100) -- (001000);  
\draw [] (0010) -- (00101); 

\draw [] (0) -- (01); 
\draw [] (01) -- (011); 
\draw [] (011) -- (0111); 
\draw [] (0111) -- (01110); 

\end{tikzpicture}
}}
		\caption{A Graph Representation of Caffeine and a path and tree decomposition of this graph. Vertices in the root bag of the tree decomposition are highlighted in green (dashed). Notice how the removal of these nodes in the original molecule separates it into two connected components, each corresponding to one of the sides of the path decomposition, and one of the highlighted subtrees in the tree decomposition.}
		\label{fig:tree_dec_caffeine}
	\end{figure}

\end{example}

\begin{lemma}[Proof in~\ref{app:proof}]\label{intseplemma}
	If $b$ is an introduce node with a single child $c$ and $X_b = X_c \cup \{v\}$, then $N(v)\cap G_b^\downarrow \subseteq X_c$. Here, $N(v)$ is the set of neighbors of $v.$
\end{lemma}

\begin{lemma}[Proof in~\ref{app:proof}]\label{joinseplemma}
	If $b$ is a join node with two children $c_1$ and $c_2,$ then in $G_b^\downarrow$ there is no edge with one endpoint in $V(G_{c_1}^\downarrow) \setminus X_b$ and the other in $V(G_{c_2}^\downarrow) \setminus X_b.$ Informally, $X_b$ is a cut that separates $V(G_{c_1}^\downarrow)$ from $V(G_{c_2}^\downarrow)$ in $G.$ See Figure~\ref{fig:tree_dec_caffeine}.
\end{lemma}

\section{Our Algorithms}\label{sec:algos} 
%

We now present our algorithms. We assume that the input contains a graph $G,$ modeling a molecule, as well as a nice decomposition of $G.$ This is without loss of generality since there are linear-time FPT algorithms parameterized by the treewidth/pathwidth to compute optimal tree/path decompositions~\cite{bodlaender1996efficient,DBLP:journals/siamcomp/Bodlaender96}. Moreover, the decompositions can also be made nice in linear time~\cite{cygan2015parameterized}. In each case, we first provide an algorithm for nice path decompositions, i.e.~handling introduce and forget nodes, and then extend it to nice tree decompositions by adding extra steps for join nodes.
Due to space restrictions and the similarity among algorithms, we have relegated most cases to~\ref{sec:remalgo}. Figures illustrating the algorithms are presented in~\ref{app:fig}.

\subsection{Counting Kekulé Structures / Perfect Matchings}\label{subsec:perfect_pathwidth}


\begin{definition}[Respectful Perfect Matchings]\label{def:respectful_perfect_matching}
Let $\mathcal{P} = \{X_1, \dots X_r\}$ be a nice path decomposition of $G.$ For each $b \in \{1, 2, \ldots, r\}$ and each $M \subseteq X_b$, we define $\RP(b,M)$ as the set of all perfect matchings $F$ in $G_b^\downarrow \setminus M$ such that each matching edge $uv \in F$ has at least one endpoint in $G_b^\downarrow \setminus X_b,$ i.e.~$u \notin X_b$ or $v \notin X_b$. We define $\pma[b, M] := |\RP(b,M)|.$
\end{definition}

\paragraph{Our Dynamic Programming Algorithm (Figures illustrating the algorithm steps are provided in \ref{appendix:figure_perfect_matching})} We note that since $X_r = \emptyset$ is the root node, $G_{r}^\downarrow = G$ and perfect matchings of $G$ are the same as respectful perfect matchings in $\RP(r,\emptyset).$ Thus $\pma [r, \emptyset]$ is the desired number of perfect matchings / Kekulé structures. We now provide a bottom-up dynamic programming approach to compute the $\pma[\cdot, \cdot]$ values. This is based on a case-work on the type of nodes:
\begin{compactitem}
\item \textbf{Leaves:} If $X_l$ is a leaf bag then $\RP(l,\emptyset)$ contains only the empty matching as $X_l = \emptyset.$ Therefore, $\pma[l,\emptyset]=1$.
\item \textbf{Introduce Nodes:} Let $b$ be a bag introducing $v$ and having a single child $c.$ We have
\[
\pma [b,M] =
\begin{cases}
\pma [c,M\setminus \{v\}] &v \in M\\
0 &v \not\in M
\end{cases}.
\]
In order to derive the above recurrence relations, we consider two possibilities for $v:$
\begin{compactenum}
\item If $v \in M$, then by Definition \ref{def:respectful_perfect_matching}, the matchings in $\RP(b,M)$ are the same as the matchings in $\RP(c, M \setminus \{v\})$. 
\item If $v \notin M$, then for every matching $F \in \RP(b,M)$, $v$ should be covered by some edge $vw \in F$. By Lemma \ref{intseplemma}, $w \in X_{c} \subset X_b$, which contradicts the requirements of Definition~\ref{def:respectful_perfect_matching}. Therefore, $\pma[b,M] = 0.$ 
\end{compactenum}


\item \textbf{Forget Nodes:} Let $b$ be a forget node with a single child $c$ and $X_b = X_c \setminus \{v\}$. We have:
\[
\pma[b,M]=
	\textstyle	\pma[c, M] + \sum_{u \in X_b\setminus M:\ uv\in E(G)}\pma[c, M \cup \{u, v\}].
\]

To see this, consider a matching $F \in \RP(b, M).$ This matching must match the forgotten vertex $v$ with another vertex $u.$ We consider two cases: (i)~the matchings for which $u \not\in X_b$ are the same as those in $\RP(c, M);$ and (ii)~the matchings $F$ for which $u \in X_b$ are counted by the sum. Here, since $v$ is matched to $u,$ neither need to be further matched in $G^\downarrow_c.$

\end{compactitem}

\begin{proposition}[Proof in~\ref{app:proof}]\label{prop:comp_pw_perfmat}
	
	Given a graph $G$ with $n$ vertices and a nice path decomposition $\mathcal{P} $ of $G$ with $O(n)$ bags and width $\pw,$ the algorithm above finds the total number of perfect matchings/Kekulé structures in time $\bigO(n\cdot \poly(\pw) \cdot2^{\pw}).$
	
\end{proposition}

\paragraph{Extension to Tree Decompositions} We now extend our algorithm above to handle tree decompositions. Assume that the input includes a nice tree decomposition $\mathcal{T} = \{T, \{X_t\}_{t \in V(T)}\}$ of the graph $G$ with width $\twi.$ As before, we perform a bottom-up dynamic programming and treat leaves and introduce/forget nodes exactly as in the previous algorithm. Computations at join nodes are performed as follows:
\begin{compactitem}
	\item \textbf{Join Nodes:} Let $b$ be a join node with children $c_1$ and $c_2.$ We have:
\[
\textstyle \pma[b,M] = \sum_{H_1 \sqcup H_2 = X_b \setminus M}\pma[c_1,M\cup H_2]\cdot \pma[c_2, M\cup H_1].
\]

where $H_1 \sqcup H_2 = X_b \setminus M$ means $H_1 \cup H_2 = X_b \setminus M$ and $H_1 \cap H_2 = \emptyset.$ 
In order to derive the above recurrence relation, let $F$ be a matching from the set $\RP(b,M)$. By Lemma~\ref{joinseplemma}, $F$ does not have any edges between $G_{c_1}^\downarrow \setminus X_b$ and $G_{c_2}^\downarrow \setminus X_b$. Therefore, it can be split into two matchings, $F_1 := E(G_{c_1}^\downarrow) \cap F$ and $F_2 := E(G_{c_2}^\downarrow) \cap F$. Let $H_1 := V(F_1) \cap X_b$ and $H_2 := V(F_2) \cap X_b$. Based on Definition~\ref{def:respectful_perfect_matching}, we have $F_1 \in \RP(c_1,M \cup H_2)$ and $F_2 \in \RP(c_2,M \cup H_1)$. If we choose $H_1$ and $H_2$ such that $H_1 \sqcup H_2 = X_b \setminus M$, then for all such matchings $F_1 \in \RP(c_1,M\cup H_2)$ and $F_2 \in \RP(c_2, M\cup H_1)$, we get a matching $ F = F_1 \cup F_2$, such that $F \in \RP(b,M).$ 



\end{compactitem}

\begin{proposition}[Proof in~\ref{app:proof}]\label{prop:comp_tw_perfmat}

	Given a graph $G$ with $n$ vertices and a nice tree decomposition $\mathcal{T}$ of $G$ with $O(n)$ bags and width $\twi,$ the algorithm above finds the total number of perfect matchings/Kekulé structures in time $\bigO(n\cdot \poly(\twi)\cdot3^{\twi})$.
\end{proposition}


\subsection{Merrifield--Simmons Index / Counting Independent Sets}\label{sec:independent_sets}


\begin{definition}[Respectful Independent Sets]\label{def:respectful_independent_sets}
Let $\mathcal{P} = \{X_1, \dots X_r\}$ be a nice path decomposition of $G.$ For each bag $b$ and each $M \subseteq X_b$, we define $\RI(b,M)$ as the set of all independent sets $I$ in a $G_b^\downarrow $ such that $I \cap X_b = M$, and $\ind[b, M]$ as the size of this set.
\end{definition}

\paragraph{Our Dynamic Programming Algorithm} Since $X_r = \emptyset$ is the root node, $G_{r}^\downarrow = G$, and all independent sets of $G$ are counted by $\ind [r, \emptyset].$ As in previous cases, our algorithm is a bottom-up dynamic programming that processes each bag according to its type:
\begin{compactitem}
\item \textbf{Leaf Nodes:} If $X_l$ is a leaf bag then $\RI(l,M)$ contains only the empty independent set as $X_l = \emptyset$. Therefore, $\ind[l,\emptyset]=1$.
\item \textbf{Introduce Nodes:} Let $b$ be an introduce bag with child $c$ and $X_b = X_c \cup \{v\}$. We have: 
\[
\ind [b,M] =
\begin{cases}
\ind [c,M] &v \notin M\\
\ind [c,M\setminus v] &v \in M \textup{ and } N(v) \cap M = \emptyset \\
0 &v \in M \textup{ and } N(v) \cap M \neq \emptyset
\end{cases}
\]
In the first case, $v$ is not meant to be in the independent set, thus we can simply look into independent sets conforming to the same $M$ in $c.$ The second case is similar, except that $v$ is in the independent set. However, note that all neighbors of $v$ in $G^\downarrow_b$ are in $X_b.$ Thus, it suffices to check whether any neighbor of $v$ is included in the independent set inside the current bag. The final case is when the set $M$ is invalid and picking neighbors.


\item \textbf{Forget Nodes:} Let $b$ be a forget node such that $X_b = X_c \setminus \{v\}$. We have $
\ind [b,M] = \ind[c,M] + \ind[c,M\cup \{v\}].$
Since $X_b \subseteq X_c,$ we know that $G^\downarrow_b = G^\downarrow_c.$ Thus, we simply need to check the two possibilities for the intersection of the independent set with $G^\downarrow_c,$ based on whether it includes or excludes $v.$

%
\end{compactitem}

\begin{proposition}\label{prop:comp_pw_ind}
	
		Given a graph $G$ with $n$ vertices and a nice path decomposition $\mathcal{P} $ of $G$ with $O(n)$ bags and width $\pw,$ the algorithm above finds the Merrifield--Simmons  index, i.e.~the total number of independent sets, in time $\bigO(n\cdot \poly(\pw) \cdot2^{\pw}).$
		
\end{proposition}

\begin{compactitem}
	\item \textbf{Join Nodes:} Let $b$ be a join node with ${c_1}$ and ${c_2}$ as its children. We have
	$
	\ind[b,M] = \ind[c_1,M] \cdot \ind[c_2,M]
	$
	This is because $M$ fixes exactly which vertices in $X_b$ are to be included in the independent set. Moreover, Lemma~\ref{joinseplemma} guarantees that only the vertices in $X_b$ are shared between $G^\downarrow_{c_1}$ and $G^\downarrow_{c_2}.$ Therefore, every independent set $I_b \in \RI(b, M)$ of $G^\downarrow_b$ is the union of a unique combination of an independent set $I_{c_1} \in \RI(c_1, M)$ of $G^\downarrow_{c_1}$ and another independent set $I_{c_2} \in \RI(c_2, M)$ of $G^\downarrow_{c_2}.$
\end{compactitem}

\begin{proposition}\label{prop:ind_tw_match}
Given a graph $G$ with $n$ vertices and a nice tree decomposition $\mathcal{T} $ of $G$ with $O(n)$ bags and width $\twi,$ the algorithm above finds the Merrifield--Simmons  index, i.e.~the total number of independent sets, in time $\bigO(n\cdot \poly(\twi) \cdot2^{\twi}).$
\end{proposition}

\todo{add figures for all cases above}

\todo{add figures for the 4 cases in all sections above}

\section{Implementation and Experimental Results}{\label{sec:Experiments}}
\label{sec:stats_molecules}
\label{sec:performance_algo}
\label{sec:implementation}

\begin{wrapfigure}{r}{5cm}
	\centering
	\vspace{-3em}
	\includegraphics[width =\textwidth]{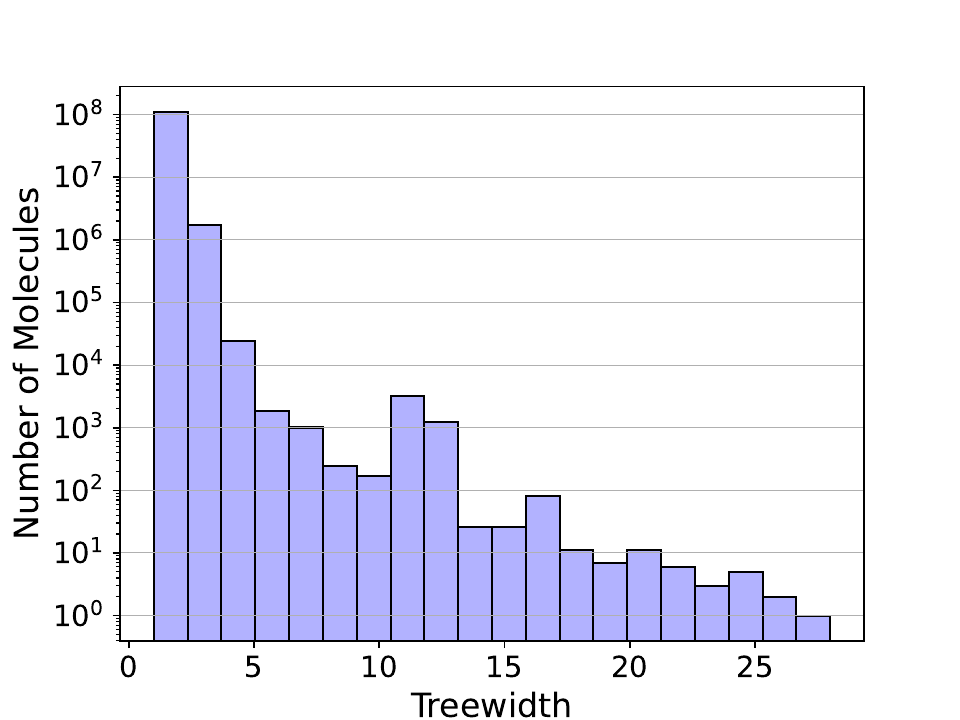}
	\vspace{-2em}
	\caption{Treewidths of PubChem Molecules.}
	\label{fig:tw_distribution}
\end{wrapfigure}

\paragraph{Benchmarks}
We conducted experiments on the entire PubChem database of chemical compunds \cite{pubchem}, containing over 113 million molecules. \ref{app:imp} has more details. 


\paragraph{Treewidth Statistics}
Figure~\ref{fig:tw_distribution} shows the treewidth distribution in the entire PubChem dataset. The $y$ axis in this figure is in logarithmic scale. Notably, more than $99.9\%$ of the compounds have a treewidth of less than 5. This holds not only for the whole database, but also for individual molecule families. See~\ref{app:fig} for more detailed tables and figures.

\begin{figure}[H]
	\resizebox{.9\linewidth}{!}{
		{\includegraphics[width=0.4\textwidth]{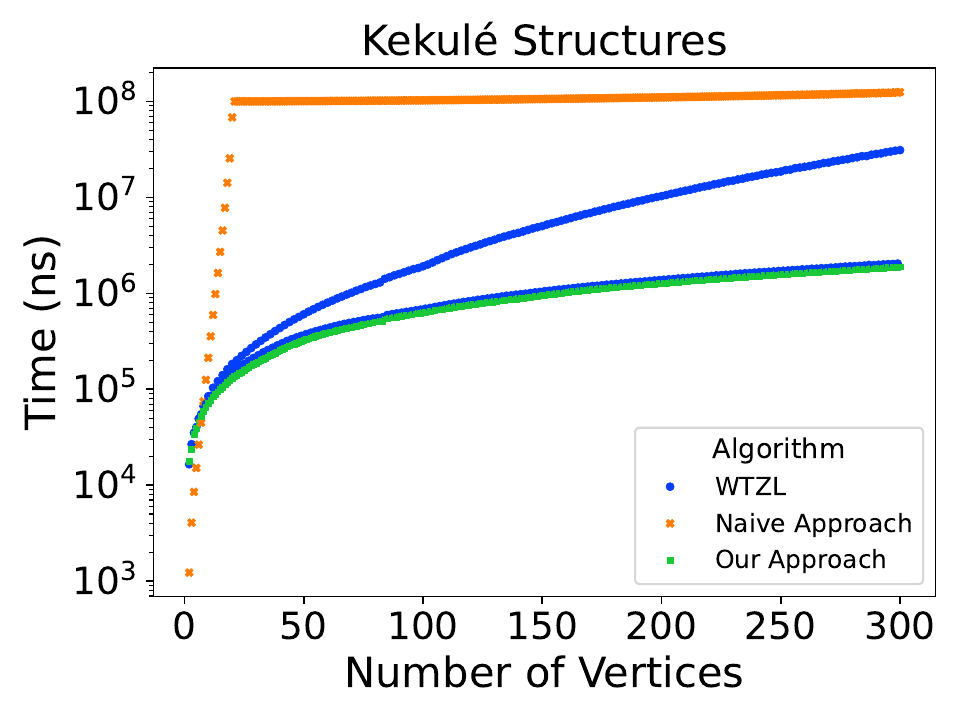}} 
		{\includegraphics[width=0.4\textwidth]{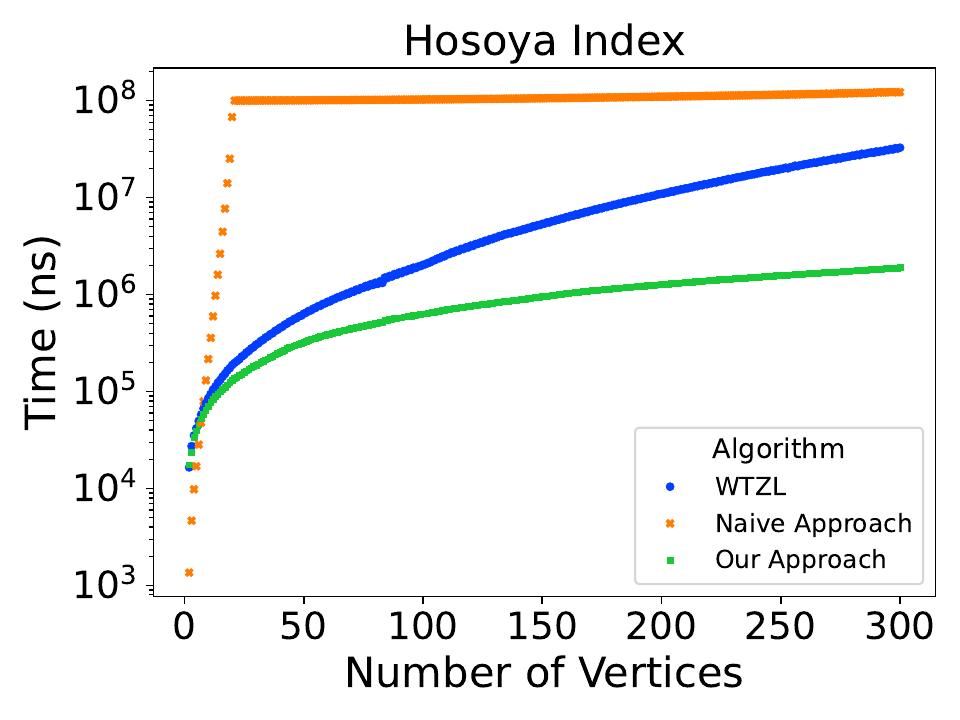}} 
		{\includegraphics[width=0.4\textwidth]{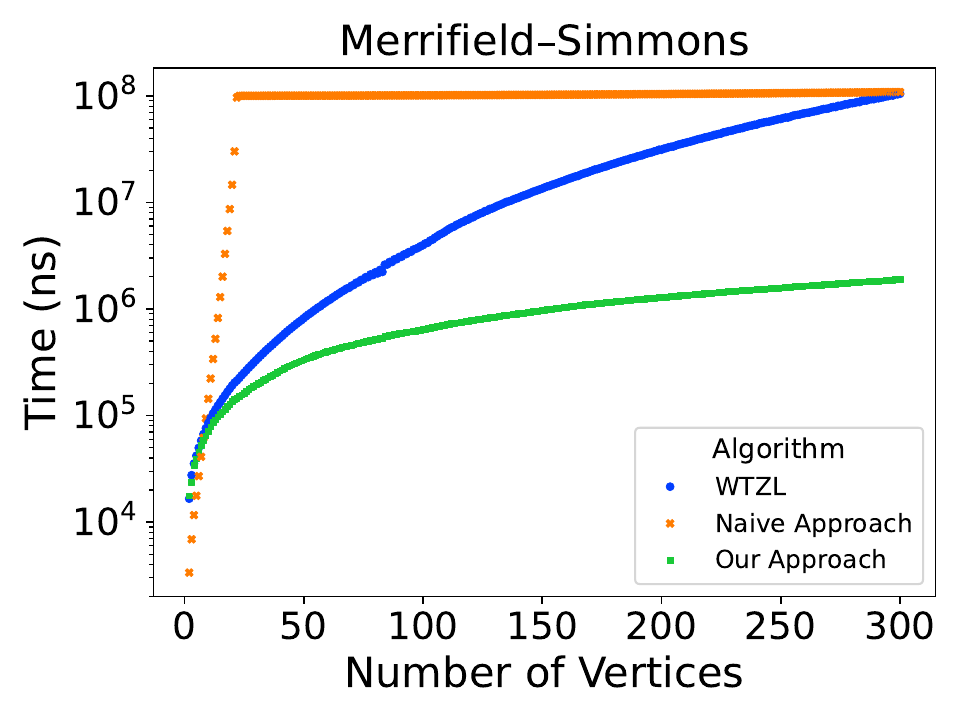}}}
	\vspace{-1em}
	\caption{Runtime Comparison of our Algorithms versus WTZL~\cite{wan2018computing} and the Naive Non-parameterized Approaches.}
	\label{fig:baseline_cmp}

\end{figure}

\vspace{-1.5em}
\begin{figure}[H]
	\resizebox{.9\linewidth}{!}{
		{\includegraphics[width=0.4\textwidth]{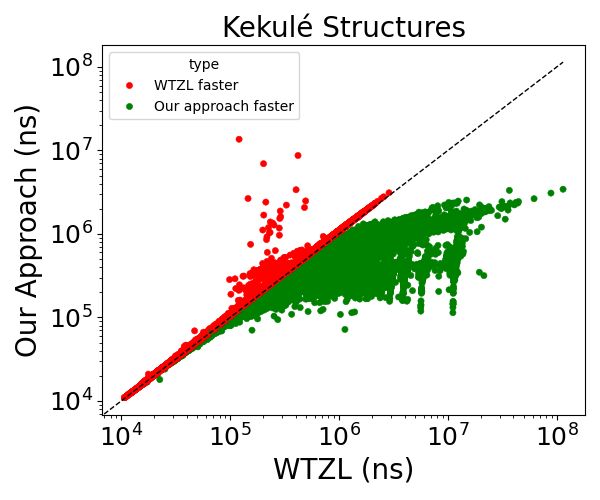}} 
		{\includegraphics[width=0.4\textwidth]{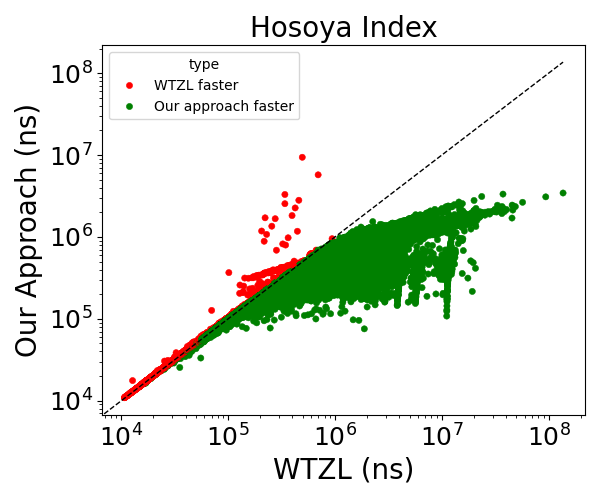}} 
		{\includegraphics[width=0.4\textwidth]{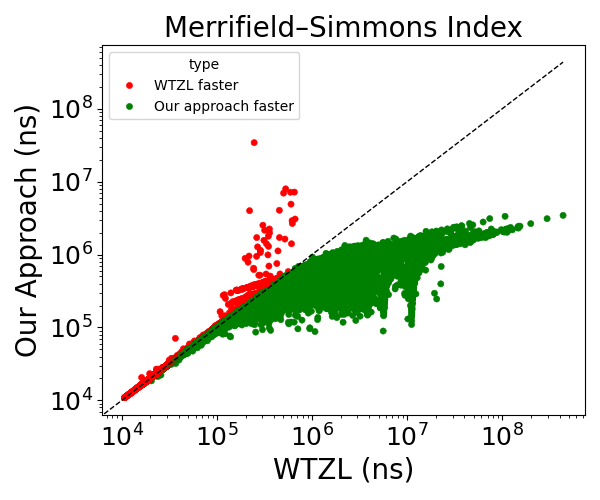}}}
	\vspace{-1em}
	\caption{Runtime Comparison of Our Approach versus WTZL \cite{wan2018computing}. 
	}
	\label{fig:wtzl_vs_our}
\end{figure}

\sloppy{
\paragraph{Runtimes}  Figure~\ref{fig:baseline_cmp} compares the runtimes of our algorithms, the WTZL algorithms~\cite{wan2018computing}, and the naive non-parameterized algorithms (\ref{app:brute}) over our benchmark set. To enable the experiment to conclude within a manageable timeframe of a few days, we enforced a time limit of 100 milliseconds per instance and report results for counting Kekul\'e structures and computing the Hosoya and Merrifield-Simmons indices. The runtimes for our approach and WTZL include the time required for computing tree decompositions. In practice, we observe that our runtime is dominated by the tree decomposition computation, rather than our own dynamic programming, whereas this is not the case for WTZL.  Note that Figure~\ref{fig:baseline_cmp} contains scatter plots, but the dots resemble lines due to the large number of benchmarks. For Kekul\'e structures, WTZL is able to conclude early over roughly half of the molecules and its runtime in these cases is close to the time required for computing tree decompositions and hence closely tracks our runtime, too. This is why Figure~\ref{fig:baseline_cmp}(a) appears to have two blue lines. We also remark that the $y$ axis in these plots is in logarithmic scale. Figure~\ref{fig:wtzl_vs_our} presents a runtime comparison between our approach and the algorithms presented in WTZL \cite{wan2018computing}. In these plots, the $x$ axis represents the time taken by the WTZL algorithms, and the $y$ axis is our time. The green points correspond to instances where our algorithm outperforms the WTZL algorithm. This accounts for {$49.3$}, {$99.04$} and {$99.20$} percent of benchmarks for Kekul\'e, Hosoya and Merrifield-Simmons, respectively. We also note that both axes in these plots are in logarithmic scale, showing that our approach is often several orders of magnitude faster in practice.
}

\bibliographystyle{elsarticle-num}
\bibliography{refs.bib}

\appendix
\newpage

\section{Proofs} \label{app:proof}

\paragraph{Proof of Lemma~\ref{intseplemma}}
		Let $T^\prime$ be a subtree of the tree $T$ rooted at $b$. Observe that the corresponding tree decomposition
		$\mathcal{T}^\prime = \{T^\prime, \{X_t\}_{t\in V(T^\prime)}\}$ is a tree decomposition of $G_b^\downarrow$. Notice that by Definition \ref{def:tree_dec}, $T^\prime_v$ forms an induced subtree of the $T^\prime$. As $v \notin X_c$ and $c$ is the only vertex in the open neighborhood of $b$, this implies that $T^\prime_v = \{b\}$.
		Let $u$ be a vertex in $N(v) \cap G_b^\downarrow$, then by Definition \ref{def:tree_dec}, if there is an edge $uv$ in $G_b^\downarrow$, then there exists a bag containing both $u$ and $v$. But, as we have just shown that the only bag containing $v$ is $X_b$. As $u$ is a vertex in $N(v)$, this implies that $u \in X_c$.

\paragraph{Proof of Lemma~\ref{joinseplemma}}
	Let $T^\prime, T^{\prime\prime}, T^{\prime\prime\prime}$ be the subtrees of $T$ rooted at $b, c_1, c_2$ respectively. Let 
	$\mathcal{T}^\prime, \mathcal{T}^{\prime\prime}, \mathcal{T}^{\prime\prime\prime}$ be the corresponding tree decompositions. We will prove the given statement by contradiction. Therefore, let us assume that $u \in V(G_{c_1}^\downarrow) \setminus X_b$, $v \in V(G_{c_2}^\downarrow) \setminus X_b$, and $uv \in E(G_b^\downarrow)$.
	As $\mathcal{T}^{\prime}$ is a tree decomposition of $G^\downarrow_b$, there exists at least one bag $X_t$, such that $t \in \mathcal{T}^{\prime}$ and $u,v \in X_t$. From our initial assumption, we know that $u, v \notin X_b$, this implies that $t \neq b$. Therefore, either $t \in T^{\prime\prime}$ or $t \in T^{\prime\prime\prime}$. Without loss of generality, let us assume that $t \in T^{\prime\prime}$. As $v \in V(G_{c_2}^\downarrow) \setminus X_b$, there exists at least one bag $X_s$ such that, $ s \in T^{\prime\prime\prime}$ and $v \in X_s$. Observe that $v$ appears in both bags $X_s$ and $X_t$. Now by Definition \ref{def:tree_dec}, we know that $T^\prime_{v}$ is a connected subtree and any path connecting $s$ and $t$ goes through $b$. This implies that $X_b \in T^\prime_v$ and $v \in X_b$. This contradicts our initial assumption that $v \notin X_b$, hence completes the proof.

\paragraph{Proof of Proposition~\ref{prop:comp_pw_perfmat}}
	At each bag, we have at most $2^{\pw}$ possibilities for $M.$ Thus, the total number of $\pma[\cdot,\cdot]$ values that need to be computed is $\bigO(n \cdot 2^\pw),$  each of which are computed in $\poly(\pw)$ time.
	
\paragraph{Proof of Proposition~\ref{prop:comp_tw_perfmat}}
	We only have to analyze the runtime at join nodes as the rest is similar to Proposition~\ref{prop:comp_pw_perfmat}. At each join node $b,$ we have to perform one step of the summation for each $M \subseteq X_b$ and each $H_1 \subseteq X_b \setminus M.$ This is because $H_2$ is uniquely determined by $H_1$ and $M.$ Consider a vertex $v \in X_b.$ The vertex $v$ is either in $M$ or in $H_1$ or in $H_2.$ Thus, it has three possibilities. Therefore, the total number of possible combinations of $M$ and $H_1$ is $3^{|X_b|} \leq 3^{\twi + 1}.$

\section{Remaining Algorithms}
\label{sec:remalgo}

\subsection{Hosoya Index / Counting Matchings}\label{sec:matchings}

 We now turn to the problem of computing the Hosoya index of a molecular graph, i.e.~finding the number of all matchings in the graph.

\begin{definition}[Respectful Matchings]\label{def:respectful_matching}
	Let $\mathcal{P} = \{X_1, \dots X_r\}$ be a nice path decomposition of $G$. For each bag $b$ and each $M \subseteq X_b$, we define $\RM(b,M)$ as the set of all matchings $F$ in  $G_b^\downarrow \setminus M$ such that (i)~each matching edge $uv \in F$ has at least one endpoint in $G_b^\downarrow \setminus X_b,$ and (ii)~each vertex in $X_b \setminus M$ is covered by $F$. We further define $\ma[b, M] := |\RM(b,M)|.$
\end{definition}

\paragraph{Our Dynamic Programming Algorithm (Figures illustrating the algorithm steps are provided in \ref{appendix:figure_matching})}
Given that $X_r = \emptyset$ is the root node, we have $G_{r}^\downarrow = G$, and all matchings of $G$ are in $\RM(r,\emptyset)$, which implies $\ma [r, \emptyset]$ is the Hosoya index of $G$. As in the previous section, we provide a bottom-up dynamic programming algorithm handling each type of node as follows:
\begin{compactitem}
	\item \textbf{Leaf Nodes:} If $X_l$ is a leaf bag then $\RM(l,M)$ contains only the empty matching as $X_l = \emptyset$. Therefore, $\ma[l,\emptyset] = 1$.
	\item \textbf{Introduce Nodes:} If $b$ introduces $v$ and has a child $c,$ we have: 
	\[
	\ma [b,M] =
	\begin{cases}
		\ma [c,M\setminus \{v\}] &v \in M\\
		0 &v \not\in M
	\end{cases}.
	\]
	In order to derive the above recurrence relation, we consider two possibilities for $v$:
	\begin{compactenum}
		\item If $v \in M$, then by Definition \ref{def:respectful_matching}, the matchings in $\RM(b,M)$ are the same as those in $\RM(c, M \setminus v)$. 
		\item If $v \notin M$, then for every matching $F \in \RP(b,M)$, $v$ should be covered by some edge $vw \in F$. By Lemma \ref{intseplemma}, $w \in X_{c} \subset X_b$, which contradicts the Definition \ref{def:respectful_matching}. Therefore, $\ma[b,M] = 0.$ for this case.
	\end{compactenum}
	
	
	\item \textbf{Forget Nodes:} Let $b$ be a forget node such that $X_b = X_c \setminus \{v\}$. We have:
	\[
	\ma[b,M]= \ma[c, M] +  \ma[c, M \cup \{v\}] \\
	+ \sum_{u \in X_b\setminus M:\ uv\in E(G)}\ma[c, M \cup \{u, v\}].
	\]
	
	This is obtained by a similar argument as in the case of perfect matchings. Every matching in $\RM(b, M)$ is in one of three cases. It either (i)~matches $v$ with a vertex outside $X_c,$ or (ii)~does not match $v$, or (iii)~matches $v$ with another vertex $u$ in $X_c.$ The matchings in (i) and (ii) are precisely those in $\RM(c, M)$ and $\RM(c, M \cup \{v\}),$ respectively, by Definition~\ref{def:respectful_matching}. For part (iii), if $v$ is matched with $u,$ then we should not match either of them to any other vertex. Thus, what remains is a matching in $\RM(c, M \cup \{u, v\}).$

\end{compactitem}


\begin{proposition}\label{prop:comp_pw_mat}
	
	Given a graph $G$ with $n$ vertices and a nice path decomposition $\mathcal{P} $ of $G$ with $O(n)$ bags and width $\pw,$ the algorithm above finds the Hosoya index, i.e.~the total number of matchings, in time $\bigO(n\cdot \poly(\pw) \cdot2^{\pw}).$
\end{proposition}
\begin{proof}
	The runtime analysis is identical to Proposition \ref{prop:comp_pw_perfmat}.
\end{proof}

\paragraph{Extension to Tree Decompositions} As in the previous section, in order to extend our algorithm to tree decompositions, we only need to specify how to handle join nodes in our bottom-up dynamic programming:
\begin{compactitem}
	\item \textbf{Join Nodes:} Let $b$ be a join node with ${c_1}$ and ${c_2}$ as its children. We then have:
	\[
	\ma[b,M] = \sum_{H_1 \sqcup H_2 = X_b \setminus M}\ma[c_1,M\cup H_2]\cdot \ma[c_2, M\cup H_1].
	\]
	
	where $H_1 \sqcup H_2 = X_b \setminus M$ means $H_1 \cup H_2 = X_b \setminus M$ and $H_1 \cap H_2 = \emptyset$. 
	In order to derive the above recurrence relation, let $F$ be a matching from the set $\RM(b,M)$. By Lemma~\ref{joinseplemma}, $F$ does not have any edges between $G_{c_1}^\downarrow \setminus X_b$ and $G_{c_2}^\downarrow \setminus X_b$. Therefore, it can be split into two matchings, $F_1 := E(G_{c_1}^\downarrow) \cap F$ and $F_2 := E(G_{c_2}^\downarrow) \cap F$. Let $H_1 := V(F_1) \cap X_b$ and $H_2 := V(F_2) \cap X_b$. Following Definition \ref{def:respectful_matching}, we have $F_1 \in \RM(c_1,M \cup H_2)$ and $F_2 \in \RM(c_2,M \cup H_1)$. If we choose $H_1$ and $H_2$ such that $H_1 \sqcup H_2 = X_b \setminus M$, then for all such matchings $F_1 \in \RM(c_1,M\cup H_2)$ and $F_2 \in \RM(c_2, M\cup H_1)$, we get a matching $ F = F_1 \cup F_2$, such that $F \in \RM(b,M)$. 

	
\end{compactitem}

\begin{proposition}\label{prop:comp_tw_match}
	
	Given a graph $G$ with $n$ vertices and a nice tree decomposition $\mathcal{T}$ of $G$ with $O(n)$ bags and width $\twi,$ the algorithm above finds the total number of perfect matchings, i.e.~the Hosoya index, in time $\bigO(n\cdot \poly(\twi)\cdot3^{\twi})$.
	
\end{proposition}
\begin{proof}
	The runtime analysis is identical to \ref{prop:comp_tw_perfmat}.
	
\end{proof}

\subsection{Counting Matchings of All Sizes}\label{sec:entropy_matchings}

In this section, given a graph $G$ and a tree/path decomposition, our goal is to find the number of matchings of size $k$ in $G$ for every $k.$
We need to only slightly adapt the dynamic programming values defined in Section~\ref{sec:matchings}, in order to count matchings of each size separately. The derivation of recurrence relations and their correctness is also very similar to Section~\ref{sec:matchings}. Finally, given this information, computing the entropy is a simple matter of applying Equation~\eqref{eq:entropy}.

\begin{definition}[Respectful Matchings of size $k$]\label{def:respectful_matching_size_k}
Given a nice path decomposition $\mathcal{P} = \{X_1, \dots X_r\}$ of $G,$ for each bag $b$ and each $M \subseteq X_b$, we define $\RM(b,k,M)$ as the set of all matchings $F$ in  $G_b^\downarrow \setminus M$ such that (i)~$\lvert F \rvert = k$; (ii)~each matching edge $uv \in F$ has at least one endpoint in $G_b^\downarrow \setminus X_b$; and (iii)~every vertex in $X_b \setminus M$ is covered by $F$. We further define $\ma[b,k,M] := |\RM(b,k,M)|.$
\end{definition}

\paragraph{Our Dynamic Programming Algorithm} Since $X_r = \emptyset$ is the root node, $G_{r}^\downarrow = G$, and $\ma [r, k,\varnothing]$ counts matchings of size $k$ in $G$. We process the bags in a bottom-up order as follows:

\begin{compactitem}
\item \textbf{Leaf Nodes:} If $X_l$ is a leaf bag then $\RM(l,k,M)$ contains only the empty matching as $X_l = \emptyset$. Therefore, $\ma[l,0,\emptyset] = 1$ and for any other $k > 0$ $\ma[l,k,\emptyset] = 0$.
\item \textbf{Introduce Nodes:} If $b$ introduces $v$ and has a single child $c$ then, we have:
\[
\ma [b,k,M] =
\begin{cases}
\ma [c,k,M\setminus v] &v \in M\\
0 &v \not\in M
\end{cases}.
\]

\item \textbf{Forget Nodes:} Let $b$ be a forget node with child $c$ and $X_b = X_c \setminus \{v\}$, then
\[
\ma[b,k,M]= \ma[c,k,M] +  \ma[c,k,M \cup \{v\}] 
+ \sum_{u \in X_b\setminus M:\ uv\in E(G)}\ma[c,k-1, M \cup \{u, v\}].
\]

\end{compactitem}

\begin{proposition}\label{prop:comp_pw_matchings_size_k} \label{prop:twe}
	
		Given a graph $G$ with $n$ vertices and a nice path decomposition $\mathcal{P} $ of $G$ with $O(n)$ bags and width $\pw,$ the algorithm above finds the number of matchings of size $k$ for every $0 \leq k \leq \frac{n}{2}$ in time $\bigO(n^2 \cdot \poly(\pw) \cdot2^{\pw}).$
	
\end{proposition}
\begin{proof}
The analysis is similar to that of Proposition~\ref{prop:comp_pw_perfmat}, except that we now have $O(n)$ times as many dynamic programming values, one for each value of $k.$
\end{proof}


\paragraph{Extension to Tree Decompositions} 
If the input contains a tree decomposition rather than a path decomposition, our algorithm proceeds in a bottom-up order and handles the join nodes as follows:
\begin{compactitem}
	\item \textbf{Join Nodes:} 
	Let $b$ be a join node with children ${c_1}$ and ${c_2}.$ We define $n_b := \lvert V(G_b^{\downarrow}) \rvert$. We define $n_{c_1}$ and $n_{c_2}$ similarly and, without loss of generality, assume that $n_{c_1} \leq n_{c_2}$\footnote{If $n_{c_1} = n_{c_2}$ we choose the lexicographically smaller bag as $c_1.$}. We call $c_1$ the \emph{light} child of $b$ and $c_2$ the \emph{heavy} child. We also note that 
	\begin{equation} \label{eq:oneandhalf} n_b = n_{c_1} + n_{c_2} - |X_b| \geq 2 \cdot n_{c_1} - |X_b|.
\end{equation}
Finally, we have:
\begin{equation} \label{eq:joinmatch}
\hspace{-1cm}\ma[b,k,M] = 
\sum_{0\leq k_{1}\leq \frac{n_{c_1}}{2}} \sum_{H_1 \sqcup H_2 = X_b \setminus M}\ma[c_1,k_{c_1},M\cup H_2]\cdot \ma[c_2,k-k_1, M\cup H_1].
\end{equation}
\end{compactitem}
This is similar to Section~\ref{sec:matchings}, except that we choose to explicitly keep track of the number of matching edges that come from $G^\downarrow_{c_1},$ i.e.~$k_1,$ and the other $k-k_1$ edges of the matching come from $G^\downarrow_{c_2}.$ Note that the first sum above is on $n_{c_1}$ where $c_1$ was the light child of $b.$

\begin{proposition}\label{prop:comp_tw_matchings_size_k}
	
		Given a graph $G$ with $n$ vertices and a nice tree decomposition $\mathcal{T}$ of $G$ with $O(n)$ bags and width $\twi,$ the algorithm above finds the number of matchings of size $k$ for every $0 \leq k \leq \frac{n}{2}$ in time $\bigO(n^2 \cdot \log n \cdot \poly(\twi) \cdot3^{\twi}).$
%
%
\end{proposition}

\begin{proof}
	All types of nodes except for join bags are covered by Proposition~\ref{prop:comp_pw_matchings_size_k}. Let $L$ be the set of all light children of join bags.
	If $c \in L$ has a corresponding subgraph with size $n_c \leq 2 \cdot (\twi+1),$ then it will contribute only $\poly(\twi)$ iterations to the first sum in Equation~\eqref{prop:comp_tw_matchings_size_k}. Let $L' = \{c \in L : n_c > 2 \cdot (\twi + 1)\}.$
	We claim that $\textstyle \sum_{c \in L'} n_c \in O(n \cdot \log n).$
	The vertex $v$ will be counted in $n_c$ if and only if there is a descendant $d$ of $c$ that introduces $v.$ 
	Consider any introduce bag $\eta$ and focus on the sequence $A_\eta$ of ancestors of $\eta$ in $T.$ The sizes of the subgraphs corresponding to the bags in this sequence are increasing as we move towards the root. If $c \in L' \cap A_\eta$ is an ancestor of $\eta$ that is also a light child of a join bag $b \in A_\eta,$ then, by Equation~\eqref{eq:oneandhalf} we have
	$
	n_b \geq 2 \cdot n_c - |X_b| \geq 2 \cdot n_c - (\twi + 1) \geq 1.5 \cdot n_c.
	$
	Thus, every time such a light ancestor is met, the size of the subgraph is multiplied by at least $1.5.$ Hence, $\eta$ can have at most $O(\log n)$ such ancestors. Since the tree decomposition has $O(n)$ bags and therefore $O(n)$ introduce bags $\eta$, each introducing a single vertex, which contributes to at most $O(\log n)$ terms of the sum, we have $\sum_{c \in L'} n_c \in O(n \cdot \log n).$ 
	Finally, the total runtime of computing the sums of the form of Equation~\eqref{eq:joinmatch} is $O\left( \left(n \cdot \log n + n \cdot \poly(\twi)\right) \cdot n \cdot 3^{\twi}\right ).$ This is because we have $O(n)$ choices for $k$ and $O(3^\twi)$ choices for $M, H_1$ and $H_2$ as argued in Proposition~\ref{prop:comp_tw_perfmat}.
\end{proof}

\subsection{Counting Independent Sets of All Sizes}
\label{sec:entropy_independent_sets}
In this section, given a graph $G$ and a nice tree/path decomposition of $G$ as input, our goal is to find the number of independent sets of size $k$ in $G$ for every possible value of $k.$ As in the previous section, simply plugging these numbers into Equation~\eqref{eq:entropy} yields the entropy.



\begin{definition}[Respectful Independent Sets of size $k$]
\label{def:respectful_independent_sets_k}
Let $\mathcal{P}$ be a nice path decomposition of $G$. For every bag $b$ and every $M \subseteq X_b$, we define $\RI(b,k,M)$ as the set of all independent sets $I$ in $G_b^\downarrow $ such that $I \cap X_b = M$ and $\lvert I \rvert=k$. We denote the size of this set as $\ind[b, k, M].$
\end{definition}

\paragraph{Our Dynamic Programming Algorithm} As in Section~\ref{sec:independent_sets}, $\ind [r, k, \emptyset]$ is the number of independent sets of size $k$ in $G$. We process our decomposition bottom-up as follows:
\begin{compactitem}
\item \textbf{Leaf Nodes:} If $X_l$ is a leaf bag then $\RI(l,k,M)$ contains only the empty independent set as $X_l = \emptyset$. Therefore, $\ind[l,0,\emptyset]=1$ and for any other $k > 0$ $\ind[l,k,\emptyset] = 0$.

\item \textbf{Introduce Nodes:} Let $X_b$ be an introduce bag with $X_b = X_c \cup \{v\}$. We have:
\[
\ind [b,k,M] =
\begin{cases}
\ind [c,k,M] &v \notin M\\
\ind [c,k-1,M\setminus v] &v \in M \textup{ and } N(v) \cap M = \varnothing \\
0 &v \in M \textup{ and } N(v) \cap M \neq \varnothing
\end{cases}.
\]
\item \textbf{Forget Nodes:} Let $X_b$ be a forget node such that $X_b = X_c \setminus \{v\}$, then
\[
\textstyle \ind [b,k,M] = \ind[c,k,M] + \ind[c,k-1,M\cup \{v\}].
\]
\end{compactitem}

\begin{proposition}\label{prop:comp_pw_ind_ent}
	
			Given a graph $G$ with $n$ vertices and a nice path decomposition $\mathcal{P} $ of $G$ with $O(n)$ bags and width $\pw,$ the algorithm above finds the number of independent sets of size $k$ for every $0 \leq k \leq n$ in time $\bigO(n^2 \cdot \poly(\pw) \cdot2^{\pw}).$
	\end{proposition}
\begin{proof}
The runtime analysis is identical to Proposition~\ref{prop:twe}.
\end{proof}

\label{sec:tw_indep_k}



\paragraph{Extension to Tree Decompositions} As in the previous cases, we only need to specify how the algorithm handles join nodes.
\begin{compactitem}
	\item \textbf{Join Nodes:} Let $b$ be a join node with light child $c_1$ and heavy child $c_2.$ We have:
		\[ \textstyle
		\ind[b, k, M] = \sum_{0\leq k_1 \leq n_{c_1}} \ind[c_1,k_1, M] \cdot \ind[c_2,k-k_1+\vert M \vert,M].
	\]
	This is because every independent set $I \in \RI(b, k, M)$ of size $k$ in the graph $G^\downarrow_b$ can be uniquely written as the union of two independent sets $I_1$ of size $k_1$ in $\RI(c_1, k_1, M)$ and $I_2$ of size $k_2$ in $\RI(c_2, k_2, M).$ Since we have $I_1 \cap I_2 = M$ and $|I_1 \cup I_2| = k,$ we must have $k_2 = k - k_1 + |M|.$
\end{compactitem}


\begin{proposition}\label{prop:comp_tw_independent_sets_size_k}
	
		Given a graph $G$ with $n$ vertices and a nice tree decomposition $\mathcal{T}$ of $G$ with $O(n)$ bags and width $\twi,$ the algorithm above finds the number of independent sets of size $k$ for every $0 \leq k \leq {n}$ in time $\bigO(n^2 \cdot \log n \cdot \poly(\twi) \cdot2^{\twi}).$
	
\end{proposition}
\begin{proof}
The runtime analysis is similar to that of Proposition~\ref{prop:comp_tw_matchings_size_k}.
\end{proof}

\section{Tables, Figures and Implementation Details} \label{app:fig} \label{app:imp} \label{app:brute}

\begin{table}[H]
	
	\resizebox{.6\linewidth}{!}{
		\begin{tabular}{lccc}
			\toprule
			& \textbf{Minimum} & \textbf{Mean} & \textbf{Maximum} \\
			\midrule
			Number of Vertices & 2  & 29.3 & 910 \\
			Number of Edges    & 1 & 31.6 & 919 \\
			Treewidth          & 1 & 1.97 & 28 \\
			\bottomrule
		\end{tabular}
	}
	
	\caption{Statistics of the PubChem Molecules.}
	\label{tab:statistics}
\end{table}

\begin{figure}[H]
	\centering
	{\includegraphics[width=0.48\textwidth]{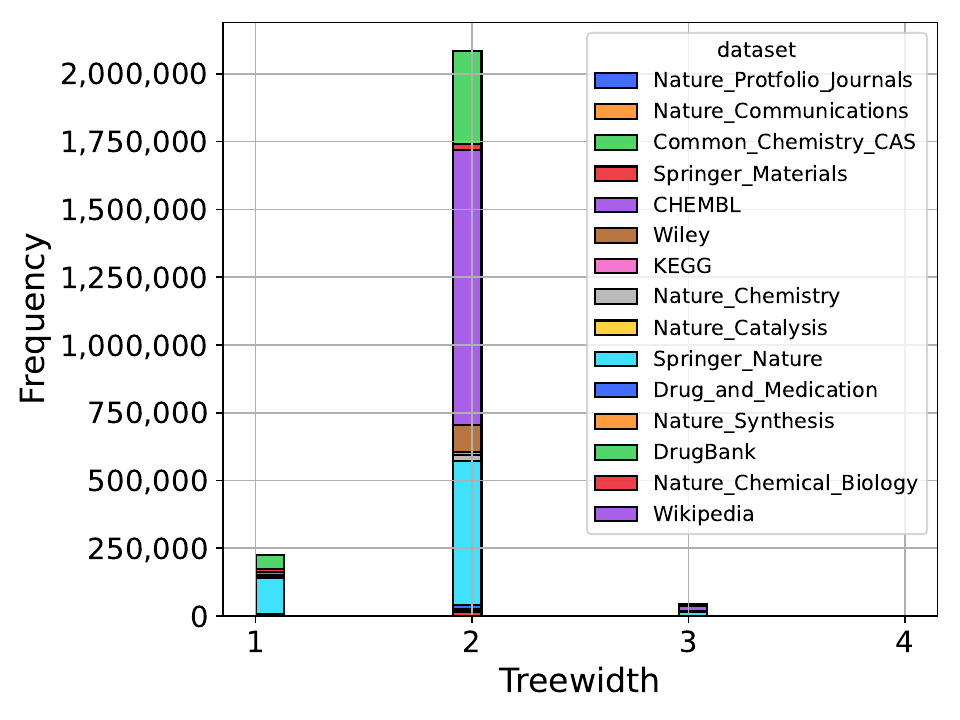}}
	{\includegraphics[width=0.48\textwidth]{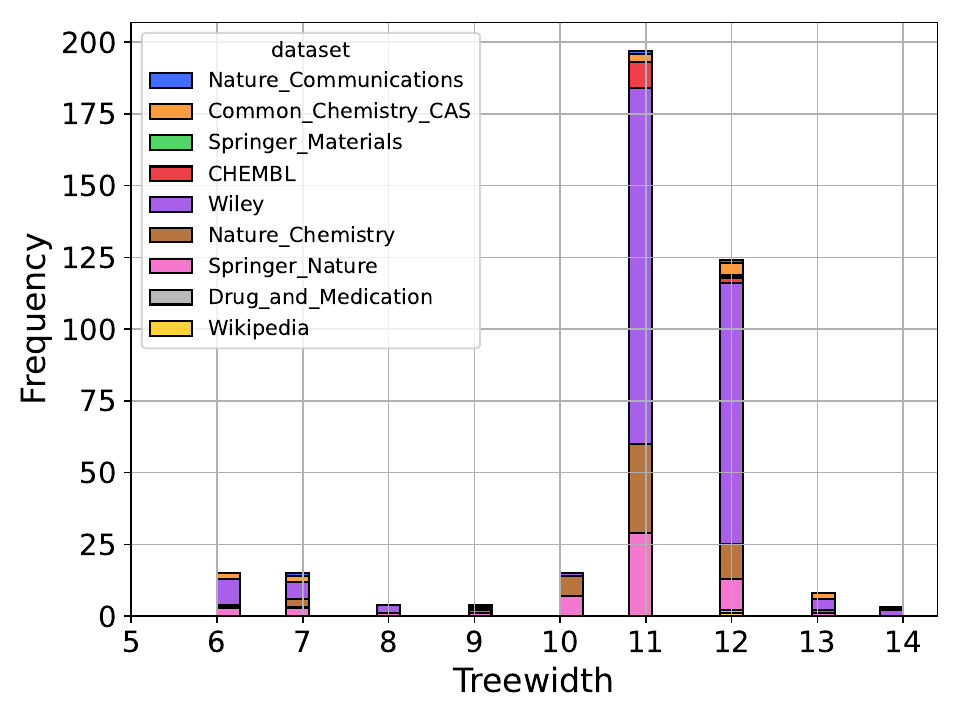}}
	\caption{Distribution of Treewidths of Molecules in Selected PubChem Datasets. Since there are many more molecules with smaller treewidths, we have broken this histogram in two parts.}
	\label{fig:tw_distribution2}
\end{figure}

\subsection{Illustrative Figures for Dynamic Programming in Perfect Matching Computation (Kekulé Structures)}
\label{appendix:figure_perfect_matching}

\begin{figure}[H]
	\floatbox[{\capbeside\thisfloatsetup{capbesideposition={right,top},capbesidewidth=7cm}}]{figure}[\FBwidth]
	{\caption{Computing values in introduce nodes. Each of the squares corresponds to a bag. In each case, bag $b$ introduces node $x_1$ and has child $c$. Nodes in red (dashed) represent $M$, i.e~nodes that are not yet ``matched''. In the left example, since $x_1\in M$, the number of respectful perfect matchings is $\pma [c,M\setminus \{x_1\}]$. In the example on the right, $x_1\not \in M$, i.e~$x_1$ should be matched with a node outside of $X_b$. This is not possible since $x_1$ has just been introduced, thus the number of respectful matchings is $0.$}\label{fig:introduce_node}}
	{\resizebox{0.35\textwidth}{!}{
\begin{tikzpicture}[line width=1pt,node distance=14mm ,main/.style = {draw, rectangle},scale=1] 
   
   \node[scale=.8] at (7,5.5){$\textup{PerfMatch}[b,\{x_1,x_2\}]$};
       \node[main] at (7,4) (9) {
   \begin{tikzpicture}[line width=1pt,node distance=14mm ,main/.style = {draw, circle},scale=1] 
   \node[main,fill=red!40!white,densely dashed] at (0,1) (x1) {\texttt{$x_1$}};
   \node[main,fill=red!40!white,densely dashed] at (1,1) (x2) {\texttt{$x_2$}};
       \node[main,color=black] at (0,0) (x3) {\texttt{$x_3$}};
       \node[main,color=black] at (1,0) (x4) {\texttt{$x_4$}};
      \draw (x1) -- (x2);
       \draw (x1) -- (x3);
      \draw (x2) -- (x3);
      \draw (x3) -- (x4);
   \end{tikzpicture}
       };

    \node[scale=.8] at (7,1.5){$\textup{PerfMatch}[c,\{x_2\}]$};
   \node[main] at (7,0) (8) {
   \begin{tikzpicture}[line width=1pt,node distance=14mm ,main/.style = {draw, circle},scale=1] 
   \node[main,fill=red!40!white,densely dashed] at (1,1) (x2) {\texttt{$x_2$}};
       \node[main,color=black] at (0,0) (x3) {\texttt{$x_3$}};
       \node[main,color=black] at (1,0) (x4) {\texttt{$x_4$}};
      \draw (x2) -- (x3);
      \draw (x3) -- (x4);
   \end{tikzpicture}
   };
   
       \draw [decorate,decoration={brace,amplitude=10}] (6,2) -- (8,2) node [black,midway,xshift=-0.6cm] {};

    \node[scale=.8] at (10.5,5.5){$\textup{PerfMatch}[b,\{x_3,x_4\}]$};
       \node[main] at (10.5,4) (9) {
   \begin{tikzpicture}[line width=1pt,node distance=14mm ,main/.style = {draw, circle},scale=1] 
   \node[main,color=black] at (0,1) (x1) {\texttt{$x_1$}};
   \node[main,color=black] at (1,1) (x2) {\texttt{$x_2$}};
       \node[main,fill=red!40!white,densely dashed] at (0,0) (x3) {\texttt{$x_3$}};
       \node[main,fill=red!40!white,densely dashed] at (1,0) (x4) {\texttt{$x_4$}};
      \draw (x1) -- (x2);
       \draw (x1) -- (x3);
      \draw (x2) -- (x3);
      \draw (x3) -- (x4);
   \end{tikzpicture}
       };

        \draw [decorate,decoration={brace,amplitude=10}] (9.5,2) -- (11.5,2) node [black,midway,xshift=-0.6cm] {};
       \node[scale=0.8] at (10.5,1.5){Cannot form};
       \node[scale=0.8] at (10.5,1.1){a respectful matching};
      
   \end{tikzpicture}
   }}
\end{figure}

\begin{figure}[H]
	\floatbox[{\capbeside\thisfloatsetup{capbesideposition={right,top},capbesidewidth=5cm}}]{figure}[\FBwidth]
	{\caption{Computing values in forget nodes. Bag $b$ forgets node $x_1$ and has child $c$. Notice that nodes $x_2$ and $x_3$ have been matched with some element not in $X_b$. Since we consider a perfect matching, $x_1$ must have been matched before being forgotten. We thus consider the cases where $x_1$ was already matched in $c$, where $x_1$ matched with $x_2$ and where $x_1$ matched with $x_3$.
		}\label{fig:forget_node}}
	{\resizebox{0.58\textwidth}{!}{
\begin{tikzpicture}[line width=1pt,node distance=14mm ,main/.style = {draw, rectangle},scale=1] 

    \node[scale=.8] at (3,6){$\textup{PerfMatch}[b,\{x_4\}]$};
       
       \node[main] at (3,4.5) (9) {
   \begin{tikzpicture}[line width=1pt,node distance=14mm ,main/.style = {draw, circle},scale=1] 
   \node[main,color=black] at (1,1) (x2) {\texttt{$x_2$}};
       \node[main,color=black] at (0,0) (x3) {\texttt{$x_3$}};
       \node[main,fill=red!40!white,densely dashed] at (1,0) (x4) {\texttt{$x_4$}};
      \draw (x2) -- (x3);
      \draw (x3) -- (x4);
   \end{tikzpicture}
       };
       
    \node[scale=.8] at (-1,1.5){$\textup{PerfMatch}[c,\{x_4\}]$};
       \node[main] at (-1,0) (8) {
   \begin{tikzpicture}[line width=1pt,node distance=14mm ,main/.style = {draw, circle},scale=1] 
   \node[main,color=black] at (0,1) (x1) {\texttt{$x_1$}};
   \node[main,color=black] at (1,1) (x2) {\texttt{$x_2$}};
       \node[main,color=black] at (0,0) (x3) {\texttt{$x_3$}};
       \node[main,fill=red!40!white,densely dashed] at (1,0) (x4) {\texttt{$x_4$}};
      \draw (x1) -- (x2);
       \draw (x1) -- (x3);
      \draw (x2) -- (x3);
      \draw (x3) -- (x4);
   \end{tikzpicture}
   };
   
    \node[scale=.8] at (3,1.5){$\textup{PerfMatch}[c,\{x_4\}\cup \{x_1,x_2\}]$};
   \node[main] at (3,0) (8) {
   \begin{tikzpicture}[line width=1pt,node distance=14mm ,main/.style = {draw, circle},scale=1] 
   \node[main,fill=red!40!white,densely dashed] at (0,1) (x1) {\texttt{$x_1$}};
   \node[main,fill=red!40!white,densely dashed] at (1,1) (x2) {\texttt{$x_2$}};
       \node[main,color=black] at (0,0) (x3) {\texttt{$x_3$}};
       \node[main,fill=red!40!white,densely dashed] at (1,0) (x4) {\texttt{$x_4$}};
      \draw [line width = 3pt, color=black]  (x1) -- (x2);
       \draw (x1) -- (x3);
      \draw (x2) -- (x3);
      \draw (x3) -- (x4);
   \end{tikzpicture}
   };
   
    \node[scale=.8] at (7,1.5){$\textup{PerfMatch}[c,\{x_4\}\cup \{x_1,x_3\}]$};
   \node[main] at (7,0) (8) {
   \begin{tikzpicture}[line width=1pt,node distance=14mm ,main/.style = {draw, circle},scale=1] 
   \node[main,fill=red!40!white,densely dashed] at (0,1) (x1) {\texttt{$x_1$}};
   \node[main,color=black] at (1,1) (x2) {\texttt{$x_2$}};
       \node[main,fill=red!40!white,densely dashed] at (0,0) (x3) {\texttt{$x_3$}};
       \node[main,fill=red!40!white,densely dashed] at (1,0) (x4) {\texttt{$x_4$}};
      \draw (x1) -- (x2);
       \draw [line width = 3pt, color=black] (x1) -- (x3);
      \draw (x2) -- (x3);
      \draw (x3) -- (x4);
   \end{tikzpicture}
   };
   
       \draw [decorate,decoration={brace,amplitude=10}] (0,2.5) -- (6,2.5) node [black,midway,xshift=-0.6cm] {};
   
      
   \end{tikzpicture}
   }}
\end{figure}

\begin{figure}[H]
	\floatbox[{\capbeside\thisfloatsetup{capbesideposition={right,top},capbesidewidth=5cm}}]{figure}[\FBwidth]
	{\caption{Computing the values at join nodes. Bag $b$ has children $c_1$ and $c_2$. Nodes $x_3$ and $x_4$ have been matched with elements not in $b$, but these elements could have been in the subtree of $c_1$ or the subtree of $c_2$. We thus iterate over all possible ways to distribute these matched nodes between $c_1$ and $c_2$. For each of these ways, we multiply the number of respectful matchings in $c_1$ and $c_2$, then, finally, we add these results together to find the answer for $b$.
		}\label{fig:join_node}}
	{\resizebox{0.6\textwidth}{!}{
\begin{tikzpicture}[line width=1pt,node distance=14mm ,main/.style = {draw, rectangle},scale=1] 

    \node[scale=.8] at (-4,1){$\textup{PerfMatch}[b,\{x_1,x_4\}]$};
       \node[main] at (-4,-0.5) (9) {
   \begin{tikzpicture}[line width=1pt,node distance=14mm ,main/.style = {draw, circle},scale=1] 
   \node[main,fill=red!40!white,densely dashed] at (0,1) (x1) {\texttt{$x_1$}};
   \node[main,color=black] at (1,1) (x2) {\texttt{$x_2$}};
       \node[main,color=black] at (0,0) (x3) {\texttt{$x_3$}};
       \node[main,fill=red!40!white,densely dashed] at (1,0) (x4) {\texttt{$x_4$}};
      \draw  (x1) -- (x2);
       \draw (x1) -- (x3);
      \draw (x2) -- (x3);
      \draw (x3) -- (x4);
   \end{tikzpicture}
       };
       
    \node[scale=.8] at (0,5.5){$\textup{PerfMatch}[c_1,\{x_1,x_4\}\cup \{x_2,x_3\}]$};
       \node[main] at (0,4) (8) {
   \begin{tikzpicture}[line width=1pt,node distance=14mm ,main/.style = {draw, circle},scale=1] 
   \node[main,fill=red!40!white,densely dashed] at (0,1) (x1) {\texttt{$x_1$}};
   \node[main,fill=red!40!white,densely dashed] at (1,1) (x2) {\texttt{$x_2$}};
       \node[main,fill=red!40!white,densely dashed] at (0,0) (x3) {\texttt{$x_3$}};
       \node[main,fill=red!40!white,densely dashed] at (1,0) (x4) {\texttt{$x_4$}};
      \draw (x1) -- (x2);
       \draw (x1) -- (x3);
      \draw (x2) -- (x3);
      \draw (x3) -- (x4);
   \end{tikzpicture}
   };
   
    \node[scale=3] at (2,4){$\cdot$};
   
    \node[scale=.8] at (4,5.5){$\textup{PerfMatch}[c_2,\{x_1,x_4\}\cup \{\}]$};
   \node[main] at (4,4) (8) {
   \begin{tikzpicture}[line width=1pt,node distance=14mm ,main/.style = {draw, circle},scale=1] 
   \node[main,fill=red!40!white,densely dashed] at (0,1) (x1) {\texttt{$x_1$}};
   \node[main,color=black] at (1,1) (x2) {\texttt{$x_2$}};
       \node[main,color=black] at (0,0) (x3) {\texttt{$x_3$}};
       \node[main,fill=red!40!white,densely dashed] at (1,0) (x4) {\texttt{$x_4$}};
      \draw (x1) -- (x2);
       \draw (x1) -- (x3);
      \draw (x2) -- (x3);
      \draw (x3) -- (x4);
   \end{tikzpicture}
   };
   
    \node[scale=.8] at (0,2.5){$\textup{PerfMatch}[c_1,\{x_1,x_4\}\cup \{x_3\}]$};
   \node[main] at (0,1) (8) {
   \begin{tikzpicture}[line width=1pt,node distance=14mm ,main/.style = {draw, circle},scale=1] 
   \node[main,fill=red!40!white,densely dashed] at (0,1) (x1) {\texttt{$x_1$}};
   \node[main,color=black] at (1,1) (x2) {\texttt{$x_2$}};
       \node[main,fill=red!40!white,densely dashed] at (0,0) (x3) {\texttt{$x_3$}};
       \node[main,fill=red!40!white,densely dashed] at (1,0) (x4) {\texttt{$x_4$}};
      \draw (x1) -- (x2);
       \draw (x1) -- (x3);
      \draw (x2) -- (x3);
      \draw (x3) -- (x4);
   \end{tikzpicture}
   };
   
    \node[scale=3] at (2,1){$\cdot$};
   
    \node[scale=.8] at (4,2.5){$\textup{PerfMatch}[c_2,\{x_1,x_4\}\cup \{x_2\}]$};
   \node[main] at (4,1) (8) {
   \begin{tikzpicture}[line width=1pt,node distance=14mm ,main/.style = {draw, circle},scale=1] 
   \node[main,fill=red!40!white,densely dashed] at (0,1) (x1) {\texttt{$x_1$}};
   \node[main,fill=red!40!white,densely dashed] at (1,1) (x2) {\texttt{$x_2$}};
       \node[main,color=black] at (0,0) (x3) {\texttt{$x_3$}};
       \node[main,fill=red!40!white,densely dashed] at (1,0) (x4) {\texttt{$x_4$}};
      \draw (x1) -- (x2);
       \draw (x1) -- (x3);
      \draw (x2) -- (x3);
      \draw (x3) -- (x4);
   \end{tikzpicture}
   };
   
    \node[scale=.8] at (0,-0.5){$\textup{PerfMatch}[c_1,\{x_1,x_4\}\cup \{x_2\}]$};
   \node[main] at (0,-2) (8) {
   \begin{tikzpicture}[line width=1pt,node distance=14mm ,main/.style = {draw, circle},scale=1] 
   \node[main,fill=red!40!white,densely dashed] at (0,1) (x1) {\texttt{$x_1$}};
   \node[main,fill=red!40!white,densely dashed] at (1,1) (x2) {\texttt{$x_2$}};
       \node[main,color=black] at (0,0) (x3) {\texttt{$x_3$}};
       \node[main,fill=red!40!white,densely dashed] at (1,0) (x4) {\texttt{$x_4$}};
      \draw (x1) -- (x2);
       \draw (x1) -- (x3);
      \draw (x2) -- (x3);
      \draw (x3) -- (x4);
   \end{tikzpicture}
   };
   
    \node[scale=3] at (2,-2){$\cdot$};
   
    \node[scale=.8] at (4,-0.5){$\textup{PerfMatch}[c_2,\{x_1,x_4\}\cup \{x_3\}]$};
   \node[main] at (4,-2) (8) {
   \begin{tikzpicture}[line width=1pt,node distance=14mm ,main/.style = {draw, circle},scale=1] 
   \node[main,fill=red!40!white,densely dashed] at (0,1) (x1) {\texttt{$x_1$}};
   \node[main,color=black] at (1,1) (x2) {\texttt{$x_2$}};
       \node[main,fill=red!40!white,densely dashed] at (0,0) (x3) {\texttt{$x_3$}};
       \node[main,fill=red!40!white,densely dashed] at (1,0) (x4) {\texttt{$x_4$}};
      \draw (x1) -- (x2);
       \draw (x1) -- (x3);
      \draw (x2) -- (x3);
      \draw (x3) -- (x4);
   \end{tikzpicture}
   };
   
    \node[scale=.8] at (0,-3.5){$\textup{PerfMatch}[c_1,\{x_1,x_4\}\cup \{\}]$};
   \node[main] at (0,-5) (8) {
   \begin{tikzpicture}[line width=1pt,node distance=14mm ,main/.style = {draw, circle},scale=1] 
   \node[main,fill=red!40!white,densely dashed] at (0,1) (x1) {\texttt{$x_1$}};
   \node[main,color=black] at (1,1) (x2) {\texttt{$x_2$}};
       \node[main,color=black] at (0,0) (x3) {\texttt{$x_3$}};
       \node[main,fill=red!40!white,densely dashed] at (1,0) (x4) {\texttt{$x_4$}};
      \draw (x1) -- (x2);
       \draw (x1) -- (x3);
      \draw (x2) -- (x3);
      \draw (x3) -- (x4);
   \end{tikzpicture}
   };
   
    \node[scale=3] at (2,-5){$\cdot$};
   
     \node[scale=.8] at (4,-3.5){$\textup{PerfMatch}[c_2,\{x_1,x_4\}\cup \{x_2,x_3\}]$};
   \node[main] at (4,-5) (8) {
   \begin{tikzpicture}[line width=1pt,node distance=14mm ,main/.style = {draw, circle},scale=1] 
   \node[main,fill=red!40!white,densely dashed] at (0,1) (x1) {\texttt{$x_1$}};
   \node[main,fill=red!40!white,densely dashed] at (1,1) (x2) {\texttt{$x_2$}};
       \node[main,fill=red!40!white,densely dashed] at (0,0) (x3) {\texttt{$x_3$}};
       \node[main,fill=red!40!white,densely dashed] at (1,0) (x4) {\texttt{$x_4$}};
      \draw (x1) -- (x2);
       \draw (x1) -- (x3);
      \draw (x2) -- (x3);
      \draw (x3) -- (x4);
   \end{tikzpicture}
   };
   
       \draw [decorate,decoration={brace,amplitude=10}] (-2,-5) -- (-2,4) node [black,midway,xshift=-0.6cm] {};
   
      
   \end{tikzpicture}}}
\end{figure}

\subsection{Illustrative Figures for Dynamic Programming in Matching Computation (Hosoya Index)}
\label{appendix:figure_matching}
\begin{figure}[H]
	\caption{Computing values in forget nodes. Cases for introduce and merge nodes are the same as to \ref{subsec:perfect_pathwidth}, represented in \ref{fig:introduce_node} and \ref{fig:join_node}, respectively. Forget nodes need to consider one more case when compared to \ref{fig:forget_node}, represented by the second case below bag $b$. This is when $x_1$ remains unmatched, since we do not need to generate a perfect matching.
	}\label{fig:forget_node_matching}
	{\resizebox{0.63\textwidth}{!}{
\begin{tikzpicture}[line width=1pt,node distance=14mm ,main/.style = {draw, rectangle},scale=1] 

    \node[scale=.8] at (3.15,6){$\textup{Match}[b,\{x_4\}]$};
       
       \node[main] at (3.15,4.5) (9) {
   \begin{tikzpicture}[line width=1pt,node distance=14mm ,main/.style = {draw, circle},scale=1] 
   \node[main,color=black] at (1,1) (x2) {\texttt{$x_2$}};
       \node[main,color=black] at (0,0) (x3) {\texttt{$x_3$}};
       \node[main,fill=blue!40!white,densely dashed] at (1,0) (x4) {\texttt{$x_4$}};
      \draw (x2) -- (x3);
      \draw (x3) -- (x4);
   \end{tikzpicture}
       };
       
    \node[scale=.8] at (-1.5,1.5){$\textup{Match}[c,\{x_4\}]$};
       \node[main] at (-1.5,0) (8) {
   \begin{tikzpicture}[line width=1pt,node distance=14mm ,main/.style = {draw, circle},scale=1] 
   \node[main,color=black] at (0,1) (x1) {\texttt{$x_1$}};
   \node[main,color=black] at (1,1) (x2) {\texttt{$x_2$}};
       \node[main,color=black] at (0,0) (x3) {\texttt{$x_3$}};
       \node[main,fill=blue!40!white,densely dashed] at (1,0) (x4) {\texttt{$x_4$}};
      \draw (x1) -- (x2);
       \draw (x1) -- (x3);
      \draw (x2) -- (x3);
      \draw (x3) -- (x4);
   \end{tikzpicture}
   };
   
    \node[scale=.8] at (1.3,1.5){$\textup{Match}[c,\{x_4\}\cup \{x_1\}]$};
   \node[main] at (1.3,0) (8) {
   \begin{tikzpicture}[line width=1pt,node distance=14mm ,main/.style = {draw, circle},scale=1] 
   \node[main,fill=blue!40!white,densely dashed] at (0,1) (x1) {\texttt{$x_1$}};
   \node[main,color=black] at (1,1) (x2) {\texttt{$x_2$}};
       \node[main] at (0,0) (x3) {\texttt{$x_3$}};
       \node[main,fill=blue!40!white,densely dashed] at (1,0) (x4) {\texttt{$x_4$}};
      \draw (x1) -- (x2);
       \draw (x1) -- (x3);
      \draw (x2) -- (x3);
      \draw (x3) -- (x4);
   \end{tikzpicture}
   };
   
    \node[scale=.8] at (4.5,1.5){$\textup{Match}[c,\{x_4\}\cup \{x_1,x_2\}]$};
   \node[main] at (4.5 ,0) (8) {
   \begin{tikzpicture}[line width=1pt,node distance=14mm ,main/.style = {draw, circle},scale=1] 
   \node[main,fill=blue!40!white,densely dashed] at (0,1) (x1) {\texttt{$x_1$}};
   \node[main,fill=blue!40!white,densely dashed] at (1,1) (x2) {\texttt{$x_2$}};
       \node[main,color=black] at (0,0) (x3) {\texttt{$x_3$}};
       \node[main,fill=blue!40!white,densely dashed] at (1,0) (x4) {\texttt{$x_4$}};
      \draw [line width = 3pt, color=black]  (x1) -- (x2);
       \draw (x1) -- (x3);
      \draw (x2) -- (x3);
      \draw (x3) -- (x4);
   \end{tikzpicture}
   };

   \node[scale=.8] at (7.8,1.5){$\textup{Match}[c,\{x_4\}\cup \{x_1,x_3\}]$};
   \node[main] at (7.8,0) (8) {
   \begin{tikzpicture}[line width=1pt,node distance=14mm ,main/.style = {draw, circle},scale=1] 
   \node[main,fill=blue!40!white,densely dashed] at (0,1) (x1) {\texttt{$x_1$}};
   \node[main,color=black] at (1,1) (x2) {\texttt{$x_2$}};
       \node[main,fill=blue!40!white,densely dashed] at (0,0) (x3) {\texttt{$x_3$}};
       \node[main,fill=blue!40!white,densely dashed] at (1,0) (x4) {\texttt{$x_4$}};
      \draw (x1) -- (x2);
       \draw [line width = 3pt, color=black] (x1) -- (x3);
      \draw (x2) -- (x3);
      \draw (x3) -- (x4);
   \end{tikzpicture}
   };
   
       \draw [decorate,decoration={brace,amplitude=10}] (-1.5,2.5) -- (7.8,2.5) node [black,midway,xshift=-0.6cm] {};
   
      
   \end{tikzpicture}
   }}
\end{figure}

\subsection{Figure Illustrating Proposition \ref{prop:comp_tw_matchings_size_k}}
\begin{figure}[H]
	\resizebox{0.32\textwidth}{!}{
\begin{tikzpicture}[node distance={17mm}, thick, main/.style = {draw, rectangle split,rectangle split parts=2}] 
\node[main] (0) []{ $n_b = 72$ \nodepart{two}$x_1,x_2,x_3$}; 
\node[main] (00) [below left of =0]{ $n_b = 40$ \nodepart{two}$x_1,x_2,x_3$};
\node[] (000) [below of =00,node distance={8mm}]{...};

\node[main,fill=cyan!20, dashed] (01) [below right of =0]{$n_b = 35$ \nodepart{two}$x_1,x_2,x_3$};
\node[] (010) [below of =01,node distance={8mm}]{...};
\node[main] (011) [below of =010,node distance={8mm}]{$n_b = 29$ \nodepart{two}$x_1,x_{10},x_{11}$};
\node[main] (0110) [below left of =011]{$n_b = 20$ \nodepart{two}$x_1,x_{10},x_{11}$};
\node[] (0110p) [below of =0110,node distance={8mm}]{...};
\node[main,fill=cyan!20, dashed] (0111) [below right of =011]{$n_b = 12$ \nodepart{two}$x_1,x_{10},x_{11}$};
\node[] (0111p) [below of =0111,node distance={8mm}]{...};
\node[main] (01110) [below of =0111p,,node distance={8mm}]{$n_b = 6$ \nodepart{two}$x_{10},\textcolor{red}{x_{15}}$};
\node[] (01110p) [below of =01110,node distance={8mm}]{...};

\draw [] (0) -- (00); 
\draw [] (00) -- (000); 

\draw [line width=3pt] (0) -- (01); 
\draw [] (01) -- (010); 
\draw [] (010) -- (011); 

\draw [line width=3pt] (011) -- (0111); 
\draw [] (011) -- (0110); 
\draw [] (0110) -- (0110p); 

\draw [] (0111) -- (0111p); 
\draw [] (0111p) -- (01110); 

\draw [] (01110) -- (01110p);

\end{tikzpicture}
}
	\caption{This figure illustrates Proposition \ref{prop:comp_tw_matchings_size_k}. The tree shown represents a nice tree decomposition. Sequences of forget and introduce nodes have been omitted by ``...''. Bags in $L$, or bags that are the light child of some join bag, have been highlighted in blue (dashed). Notice that each of these bags has a parent $b$ with weight $n_b$ at least $1.5$ times the weight of its lightest child. Thus, a node, such as $x_{15}$, highlighted in red, can have at most $\log n$ light ancestors.
	}
	\label{fig:log_tree}
	
\end{figure}

\paragraph{Implementation}
We implemented our algorithms, as well as those of~\cite{wan2018computing} and the naive brute-force approaches (See \ref{app:brute}), in \texttt{C++}. We used FlowCutter~\cite{strasser2017computing} to obtain the tree decompositions. The implementation will be published as free and open-source software in the public domain and with no copyright.

\paragraph{Machine}
All experimental results were obtained on an Intel Xeon Gold 5115 Machine (2.40GHz, 16 cores) with 64 GB of RAM, running Ubuntu 22.04. The computations took approximately three days to complete.


\paragraph{Naive Baseline Algorithms} We also provide naive non-parameterized algorithms for the problems considered in this work. These are the non-parameterized algorithms used in our experimental results. The parameterized baselines were taken from~\cite{wan2018computing}. We denote the numbers of perfect matchings, matchings and indpendent sets of $G$ by $\pma(G), \ma(G)$ and $\ind(G)$ respectively. We obtain all of these values by dynamic programming using the following recurrences.



	\[
	\pma(G) =
	\begin{cases}
		\pma(G - \{u, v\}) + 
		\pma(G - uv) & uv \in E(G)  \\
		1 & E(G) = V(G) = \emptyset \\
		0 & E(G) = \emptyset, V(G) \neq \emptyset
	\end{cases}
	\]

	\[
	\ma(G) =
	\begin{cases}
			\ma(G - \{u, v\}) + \ma(G - uv) & uv \in E(G) \\
		1 & E(G) = \emptyset\\
	\end{cases}
	\]

	\[
	\ind(G) =
	\begin{cases}
		\ind(G - N[v]) + \ind(G - v) & E(G) \neq \emptyset \\
		2^{\abs{V(G)}} & E(G) = \emptyset
	\end{cases}
	\]

\newpage







\end{document}